\documentclass[]{aa}

\usepackage[varg]{txfonts}
\usepackage[english]{babel}
\usepackage{graphicx}
\usepackage{supertabular}
\usepackage{longtable}
\usepackage{lscape}
\usepackage[]{natbib}
\bibpunct{(}{)}{;}{a}{}{,}
\usepackage{verbatim}
\bibpunct{(}{)}{;}{a}{}{,}
\usepackage{ulem}
\usepackage{xcolor}
\usepackage{version}
\usepackage{mathtools}
\usepackage{comment}

\def\Ia{SN~Ia}

\def\lya{Ly$\alpha~$}
\def\lyb{Ly$\beta~$}
\def\lyalya{Ly$\alpha$(Ly$\alpha$)~}

\def\lyalyab{Ly$\alpha$(Ly$\alpha+$Ly$\beta$)~}
\def\alllya{\lyalya $\times$ \lyalyab~}
\def\autolya{Ly$\alpha$(Ly$\alpha$) $\times$ Ly$\alpha$(Ly$\alpha$)~}
\def\crosslya{Ly$\alpha$(Ly$\alpha$) $\times$ Ly$\alpha$(Ly$\beta$)~}
\def\crosslyb{Ly$\alpha$(Ly$\alpha$) $\times$ Ly$\beta$(Ly$\beta$)~}
\def\hMpc{$h^{-1}\rm{Mpc}$}
\DeclareMathOperator{\sinc}{sinc}

\newcommand{\lrf}{\lambda_{\rm RF}}
\newcommand{\kpar}{k_{\parallel}}
\newcommand{\apar}{\alpha_{\parallel}}
\newcommand{\aperp}{\alpha_{\perp}}
\newcommand{\rpar}{r_{\parallel}}
\newcommand{\rperp}{r_{\perp}}
\newcommand{\blya}{b_{\rm Ly\alpha}}
\newcommand{\betalya}{\beta_{\rm Ly\alpha}}

\newcommand{\betalyb}{\beta_{\rm Ly\beta}}
\newcommand{\dlya}{d_{\rm Ly\alpha}}
\newcommand{\bhcd}{b_{\rm HCD}}
\newcommand{\betahcd}{\beta_{\rm HCD}}
\newcommand{\Fhcd}{F_{\rm HCD}}
\newcommand{\Lhcd}{L_{\rm HCD}}
\newcommand{\imin}{i_{\rm min}}
\newcommand{\imax}{i_{\rm max}}

\newcommand{\jmax}{j_{\rm max}}
\newcommand{\xioned}{\xi_{\rm 1d}}

\author{Victoria~de~Sainte~Agathe\inst{\ref{inst1}}\thanks{Contact \email{victoria.de.sainte.agathe@lpnhe.in2p3.fr}}
\and Christophe~Balland\inst{\ref{inst1}}
\and Hélion~du~Mas~des~Bourboux\inst{\ref{inst2}}
\and Nicol{\'a}s~G.~Busca\inst{\ref{inst1}}
\and Michael~Blomqvist\inst{\ref{inst3}}
\and Julien~Guy\inst{\ref{inst4}}
\and James~Rich\inst{\ref{inst5}}
\and Andreu Font-Ribera\inst{\ref{inst6}}
\and Matthew M. Pieri\inst{\ref{inst3}}
\and Julian~E.~Bautista\inst{\ref{inst7}}
\and Kyle~Dawson\inst{\ref{inst2}}
\and Jean-Marc~Le Goff\inst{\ref{inst5}}
\and Axel de la Macorra\inst{\ref{inst8}}
\and Nathalie~Palanque-Delabrouille\inst{\ref{inst5}}
\and Will~J.~Percival\inst{\ref{inst9}}
\and Ignasi~P{\'e}rez-R{\`a}fols\inst{\ref{inst3}}
\and Donald~P.~Schneider\inst{\ref{inst10},\ref{inst11}}
\and An\v{z}e~Slosar\inst{\ref{inst12}}
\and Christophe~Y{\`e}che\inst{\ref{inst5}}
}

\institute{Sorbonne Universit\'e, Universit\'e Paris Diderot, CNRS/IN2P3, Laboratoire de Physique Nucl\'eaire et de Hautes Energies, LPNHE, 4 Place Jussieu, F-75252 Paris, France\label{inst1}
\and Department of Physics and Astronomy, University of Utah, 115 S. 1400 E., Salt Lake City, UT 84112, U.S.A.\label{inst2}
\and Aix Marseille Univ, CNRS, CNES, LAM, Marseille, France\label{inst3}
\and Lawrence Berkeley National Laboratory, 1 Cyclotron Road, Berkeley, CA 94720, U.S.A.\label{inst4}
\and IRFU, CEA, Universit{\'e} Paris-Saclay, F-91191 Gif-sur-Yvette, France\label{inst5}
\and University College London, Gower St, Kings Cross, London WC1E6BT\label{inst6}
\and Institute of Cosmology \& Gravitation, University of Portsmouth, Dennis Sciama Building, Portsmouth, PO1 3FX, UK\label{inst7}
\and Instituto de Astronomıa, Universidad Nacional Aut\'{o}noma de M\'{e}xico, A.P. 70-264, 04510, M\'{e}xico, D.F., M\'{e}xico\label{inst8}
\and Department of Physics and Astronomy, University of Waterloo, 200 University Ave. W., Waterloo ON N2L 3G1, Canada\label{inst9}
\and Department of Astronomy and Astrophysics, The Pennsylvania State University, University Park, PA 16802\label{inst10}
\and Institute for Gravitation and the Cosmos, The Pennsylvania State University, University Park, PA 16802\label{inst11}
\and Brookhaven National Laboratory, 2 Center Road,  Upton, NY 11973, USA\label{inst12}
}

\usepackage{twoopt}
\makeatletter
\begin{document}

\title{Baryon acoustic oscillations at $z = 2.34$ from the correlations of Ly$\alpha$  absorption in eBOSS DR14}
\date{Received / Accepted}
\abstract{We measure the imprint of primordial baryon acoustic
  oscillations (BAO) in the
  correlation function of Ly$\alpha$~absorption in quasar spectra 
  from the
  Baryon Oscillation Spectroscopic Survey (BOSS) and the
  extended BOSS (eBOSS) in Data Release 14 (DR14) of the
  Sloan Digital Sky Survey
  (SDSS)-IV.
  In addition to  179,965 spectra with absorption in the
  Lyman-$\alpha$ (Ly$\alpha$) region, we use, for the first time,
  Ly$\alpha$ absorption in the Lyman-$\beta$ region of 
  56,154 spectra.
  We measure the Hubble distance, $D_H$,  and the
  comoving angular diameter distance, $D_M$, 
  relative to the sound horizon at the drag epoch $r_d$ at an
  effective redshift $z=2.34$.
  Using a physical model of  the correlation function outside the BAO peak,
  we find
  $D_H(2.34)/r_d=8.86\pm 0.29$ and  $D_M(2.34)/r_d=37.41\pm 1.86$,
  within 1$\sigma$ from the flat-$\Lambda$CDM model  consistent with
  CMB anisotropy measurements.
  With the addition of
  polynomial ``broadband'' terms, the results
  remain within one standard deviation of the CMB-inspired model.
  Combined with the quasar-Ly$\alpha$ cross-correlation measurement
  presented in a companion paper \citep{Blomqvist19},
  the BAO measurements at $z=2.35$ are within 1.7$\sigma$ of the
  predictions of this model.}


\titlerunning{BAO from correlations of Ly$\alpha$ absorption in eBOSS DR14}

\maketitle

\section{Introduction}

Since the first observations of the imprint of
primordial baryonic acoustic oscillations (BAO)
as a peak in the galaxy correlation function \citep{Eiseintein05}
or as a periodic modulation in the corresponding power spectrum
\citep{Cole05},
the BAO signal
has led to significant constraints on  cosmological parameters.
The BAO peak in 
the radial direction at a redshift $z$ yields
${D_H(z)/r_d=c/(r_d H(z))}$,
where $H(z)$ is the Hubble parameter and $r_d$ is the sound horizon
at the drag epoch \citep{EisensteinHu1998}.
The transverse measurement constrains the quantity
${D_M(z)/r_d=(1+z)D_A(z)/r_d}$, where $D_A(z)$ is the angular diameter distance.
Because of its sensitivity to both the distance and the expansion rate,
the ensemble of BAO  measurements
yield tight constraints on $\Lambda$CDM parameters \citep{Aubourg15}
even without the use of Cosmic Microwave Background (CMB) data.

\begin{figure*}
	\includegraphics[width=\textwidth]{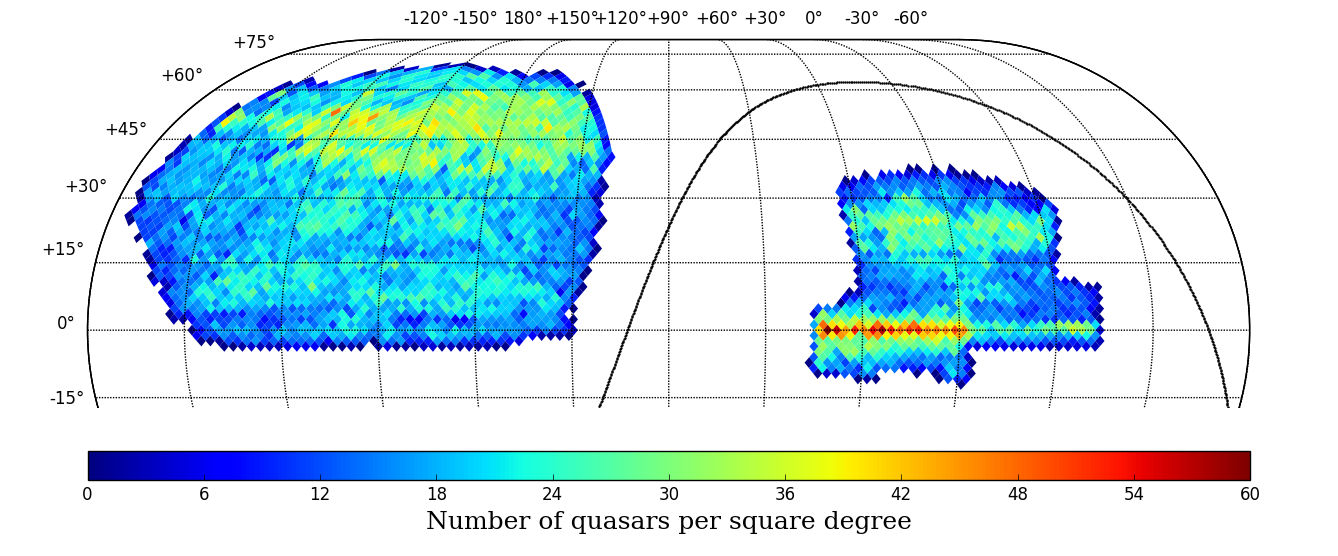}
	\caption{\label{fig:qso_map} Sky distribution of the 216,163 quasars with redshift in the [2.0,3.5] range in the DR14 footprint of the BOSS and eBOSS surveys.
 The high-density regions are the eBOSS and SEQUELS observations (for the highest declinations in the two Galactic caps) and SDSS-stripe 82 (on the celestial equator in the south galactic cap).
}
\end{figure*}

Most BAO measurements have employed discrete objects like galaxies
\citep{Percival10, Reid10, Beutler11, Blake11, Anderson12, Anderson14a, Anderson14b, Ross15, Alam17, Bautista18}
or quasars
\citep{Ata18,GilMarin18,Hou18,Zarrouk18}.
An alternative tracer of the density is
the intergalactic medium (IGM), itself
traced by  \lya absorption in quasar spectra.
Such measurements at $z\approx 2.4$ were  suggested by
\citet{McDonald03} and \citet{McDonald07}.
The first detections of a BAO peak in the Ly$\alpha$
auto-correlation function
\citep{Busca13,Slosar13}
used data from the  Baryon Oscillation Spectroscopic Survey (BOSS)
in the Sloan Digital Sky Survey (SDSS) data-release 9 (DR9),
at an effective redshift of $z=2.3$.
\citet{Delubac15}, using BOSS in SDSS-DR11, confirmed the detection of a
BAO acoustic peak in the Ly$\alpha$ auto-correlation function
at the 5$\sigma$ level.
Most recently,
\citet{Bautista17} (B17 hereafter)
used Ly$\alpha$ forests from BOSS DR12 data and provided a measurement of $D_H/r_d$ at 3.4\% precision level (or of the optimal combination $D_H^{0.7}D_M^{0.3}/r_d$ at the 2.5\% level).
The results were within 1$\sigma$ of the prediction of the flat $\Lambda$CDM model favored by CMB anisotropies \citep{Planck16cosmo}. 
However, when
combined with the BAO imprint on the cross-correlation of the Ly$\alpha$ forest
with BOSS DR12 quasars \citep{Bourboux17}, the values of
$D_H/r_d$ and $D_M/r_d$ at $z\sim2.3$ differ by $2.3\sigma$ from this model.
This mild tension was already present in the combined
constraints of the cross-correlation
measurement of \citet{Font-Ribera14} and the auto-correlation of
\citet{Delubac15}.

In the present paper,
  we use quasar spectra from the BOSS survey and from its extended version eBOSS in the
  SDSS DR14 to study BAO in the \lya auto-correlation function.
  The quasar-\lya cross-correlation is studied in a  companion
  paper \citep{Blomqvist19}
As in previous measurements, we use \lya absorption in 
the ``\lya region'' of quasar spectra, i.e., quasar rest-frame wavelengths
in the range $104<\lrf<120$~nm.
We call the auto-correlation function using only this region the
\autolya correlation\footnote{We use
  the notation ``absorption(spectral region)'' to distinguish the nature of absorption from the wavelength interval where the absorption is observed.
  Hence, the notation
  \autolya
  denotes Ly$\alpha$ absorption in Ly$\alpha$ regions correlated with
  Ly$\alpha$ absorption in Ly$\alpha$ regions.
}.
To increase the statistical power
we also include Ly$\alpha$ absorption in the Ly$\beta$ regions of quasars,
$97.4<\lrf<102$~nm, 
correlated with the Ly$\alpha$ absorption in Ly$\alpha$
regions, i.e.,
the Ly$\alpha$(Ly$\alpha$)xLy$\alpha$(Ly$\beta$) correlation function.
The Ly$\beta$ region was previously used
by \citet{Irsic13} to investigate the flux transmission power spectrum within
individual spectra.

  Besides the use of the \lyb region, the analysis
  presented here differs in several ways from that of \citet{Bautista17}
  based on DR12 data.
  For each quasar, we now use all observations instead of just the
  best one. We analyze $\sim 15$\% more Ly$\alpha$ regions.
  We have refined the modelling of the weights  (\S \ref{sec:weights}),
  the way we take into account the effect of unmasked High Column Density
  (HCD) systems, and the modeling of nonlinearities in the
  power spectrum (\S \ref{subsec:baseline}).
  We have not developed new mock spectra beyond those used in the DR12 analysis
  though this is being done
  for the final eBOSS analysis.

The layout of the paper is the following. In \S \ref{sec:data},
we present the Ly$\alpha$ and Ly$\beta$ spectral region samples used
in the present study.
We compute correlation function of \lya absorption
for the DR14 data in \S \ref{sec:lya_cf} and present our physical model
for this function in \S \ref{sec:model_lya}.
The results of fitting the data are presented and discussed in \S \ref{sec:fit}.
We draw cosmological conclusions in \S \ref{sec:cosmo} and
summarize our results in \S \ref{sec:conclu}.

The computations of the correlation functions presented in this paper have been performed using a dedicated software package, Package for Igm Cosmological-Correlations Analyses ({\tt picca}), developed by our team\footnote{Available at {\tt https://github.com/igmhub/picca}.}.

\section{Data sample and reduction }
\label{sec:data}

\begin{figure}
	\includegraphics[width=\columnwidth]{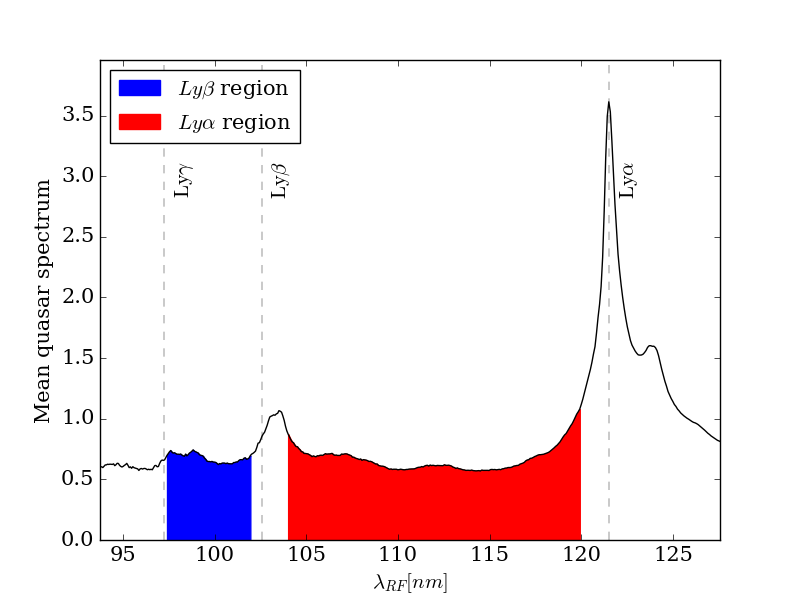}
	\caption{The Ly$\alpha$ and Ly$\beta$ spectral regions
          defined in Table \ref{table:forests}.
        }
	\label{fig:forests}
\end{figure}

\begin{figure}
	\includegraphics[width=\columnwidth]{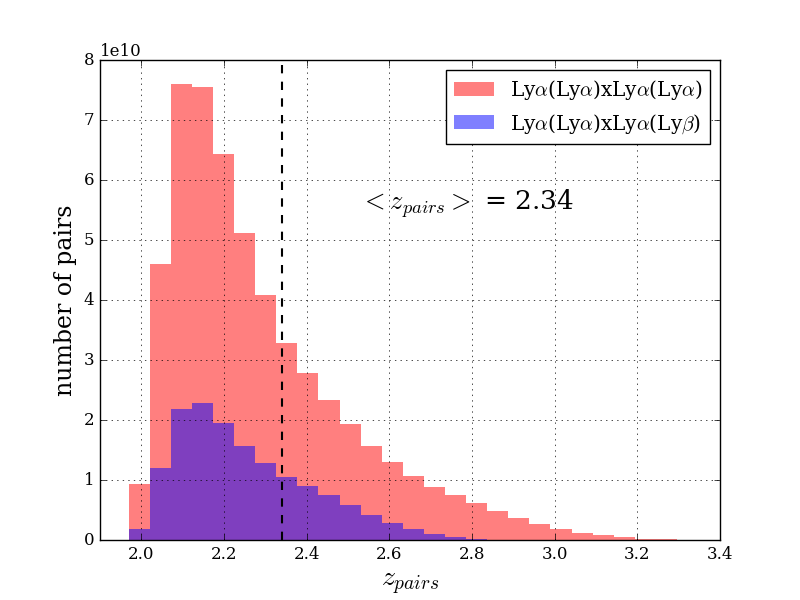}
	\caption{Weighted distribution of the redshift of pairs used to
          measure the \autolya and \crosslya correlation functions.
          The mean redshift of the combined sample is
          $\langle z_{pairs}\rangle=2.34$.
        }
	\label{fig:redshift_dist}
\end{figure}

The extended Baryon Oscillation Spectroscopic Survey (eBOSS; \citealt{Dawson16}) is the extension of the BOSS experiment \citep{Dawson13} which aims to measure cosmology with BAO using optical spectra from quasars, emission line galaxies and luminous red galaxies. It is one of the four projects of the fourth stage of the Sloan Digital Sky Survey (SDSS-IV; \citealt{Blanton17}). \\

The optical spectra are collected from 1000 fibers, attached to the
focal plane of a 2.5 m telescope in Apache Point Observatory \citep{Gunn06}, by two spectrographs in the wavelength range $[360,1000]$ nm \citep{Smee13}. The spectral resolution of the spectrographs is $\approx 2000$.\\

In this paper, we use the forests of the high-redshift quasar sample from the SDSS Data Release 14 (DR14; \citealt{Abolfathi17}). This sample contains the first two years of eBOSS data and the five years of BOSS observations reprocessed using the eBOSS pipeline. It also includes data from the ancillary programs Time-Domain Spectroscopic Survey (TDSS) and SPectroscopic IDentification of ERosita Sources (SPIDERS). 
The quasar target selection is presented in \citet{Myers15}. Note that eBOSS also targets quasars at low redshifts (where the Ly$\alpha$ region is not
observable) to be used in other programs
\citep{Ata18, Wang18, Blomqvist18, Zhao19,Helion19}.

The automated data reduction is organized in two steps (\citealt{Dawson16}). The pipeline initially extracts the two-dimensional raw data into a one-dimensional 
flux-calibrated spectrum. During this procedure, the spectra are wavelength and flux calibrated and the individual exposures of one object are coadded into a rebinned spectrum with $\Delta \log (\lambda) = 10^{-4}$. 
The spectra are then classified as STAR or GALAXY or QSO, and their redshift is estimated.
Objects that cannot be automatically classified are visually inspected \citep{Paris17} and a quasar catalog is produced, which contains 526,356 quasar spectra with redshift $0<z<7$. Among these objects, 144,046 were not in DR12. The coverage footprint of DR14 quasars is presented in Fig.\ref{fig:qso_map} \\
\begin{table}
  \caption{\label{table:forests} Definitions of the Ly$\alpha$ and Ly$\beta$ regions in terms of restframe wavelength range.
    Also listed are the allowed  observer frame wavelength ranges,
    the corresponding quasar redshift ranges and the
    number of forests available in our sample.} 
	\begin{tabular}{ccccc}
	\hline 
	\hline 
	 Regions & $\lambda_{\rm{RF}}[\rm{nm}]$ & $\lambda_{\rm{obs}}~[\rm{nm}]$ & $z_q$  & $\#$ forests \\ 
	\hline 
    Ly$\beta$ & [97.4,102] & [360,459] & [2.53,3.5] & 56,154\\
    Ly$\alpha$ & [104,120] & [360,540] & [2.0,3.5] & 179,965\\ 
	\hline 
	\end{tabular} 
\end{table}

In this study, we examine both \lya and \lyb regions (see Fig. \ref{fig:forests}). The \lya region in the quasar spectrum lies between the \lya and the \lyb emission peaks.
We limit its coverage to the rest-frame wavelength range [104,120] nm in order not to include the emission peaks, whose shape depends on the environment of the quasar. This approach minimizes the variance of the flux-transmission field defined in section \ref{sec:deltas}. Similarly, we define the \lyb region as the rest-frame wavelength range [97.4,102] nm (Table \ref{table:forests}, Fig. \ref{fig:forests}).

In the DR14 quasar catalog, selecting quasar redshifts in the range [2.0, 3.5] yields to 216,162 spectra containing, at least partially, the \lya region, and selecting quasar redshifts in the range $z_q \in [2.53, 3.5]$ yields 86,245 spectra containing the \lyb region. We choose $z_q=3.5$ as an upper limit, as beyond this redshift the quasar density is insufficient to measure correlations and
the rate of redshift misidentification is large \citep{Busca18}.
The requirement that the observed wavelength must be greater than 360 nm is due to the low CCD response and atmospheric transmission in the UV region.

In order to mask damped \lya systems (DLA),
we use the updated DR14 DLA catalog of \citet{Noterdaeme09, Noterdaeme12},
which contains 34,541 DLA in  27,212 forests.
The absorption of the identified DLAs are modeled with a Voigt profile
and the regions with more than 20\% of absorbed flux are masked.
For the \lyb regions, we apply this procedure both for \lya and \lyb
strong absorbers. 
We also mask the sky emission and absorption lines listed on the
SDSS website\footnote{http://classic.sdss.org/dr6/algorithms/linestable.html}.
The Broad Absorption Line (BAL) quasars are automatically identified
\citep{Paris17} and excluded from the data,
leaving a sample of 201,286 objects for the Ly$\alpha$ regions and
80,443 for the Ly$\beta$ regions. 

For the determination of the correlation function, we divide spectra
into ``analysis pixels'' 
that are the inverse-variance-weighted flux average
over three adjacent pipeline pixels.
Throughout the rest of this paper, ``pixel'' refers to analysis
pixels unless otherwise stated.
Spectral regions with less than 50 such pixels in regions or which have failed the continuum-fitting procedure (Sec. \ref{sec:deltas}) are discarded. These selection criteria produce 179,965 Ly$\alpha$ regions (compared to  157,783
in B17) and 56,154 Ly$\beta$ regions (see Table \ref{table:forests}).

The analysis procedure described in the next section
assigns redshifts to the observed pixel wavelengths by assuming
that flux decrements,  in both the \lya and \lyb regions,
are due to \lya absorption.
The effect of non-\lya absorption is
taken into account in the correlation-function
model presented in \S \ref{sec:model_lya}.
The weighted distribution of the redshifts of pairs of pixels used to measure
\autolya and \crosslya correlations are presented in
Fig. \ref{fig:redshift_dist}.
The mean redshift of the combined set of pixel pairs is 2.34.

\begin{table}
  \caption{\label{tab:fidcosm}Parameters of the ``Pl2015 model'', i.e.
    the flat $\Lambda$CDM model of \citet{Planck16cosmo} that we
    use here to transform
    redshifts and angular separations  into radial and transverse separations.} 
  \centering
  \begin{tabular}{ll}
    \hline 
    \hline 
    Parameters & Values \\ 
    \hline 
    $\Omega_Mh^2$ & 0.1426\\
    $=\Omega_Ch^2+\Omega_Bh^2+\Omega_{\nu}h^2$ & 0.1197+ 0.02222+ 0.0006\\
    $h$ & 0.6731\\
    $N_{\nu}$, $\sigma_8$, $n_s$& 3,\hspace*{1mm} 0.8298,\hspace*{1mm} 0.9655\\
    \hline 
    $\Omega_m$ & 0.3147 \\
    $r_d$ & 147.33~Mpc \hspace*{1mm}(99.17~\hMpc) \\
    $D_H(2.34)/r_d$,\hspace*{1mm} $D_M(2.34)/r_d$  &8.581,\hspace*{2mm} 39.26 \\
    \hline
  \end{tabular} 
\end{table}

\section{Computing the Ly$\alpha$ correlation function from the data}
\label{sec:lya_cf}

This section describes, first,  the measurement of the
flux-transmission field and then its correlation function
and associated covariance matrix.

\begin{figure}[t]
	\includegraphics[width=\columnwidth]{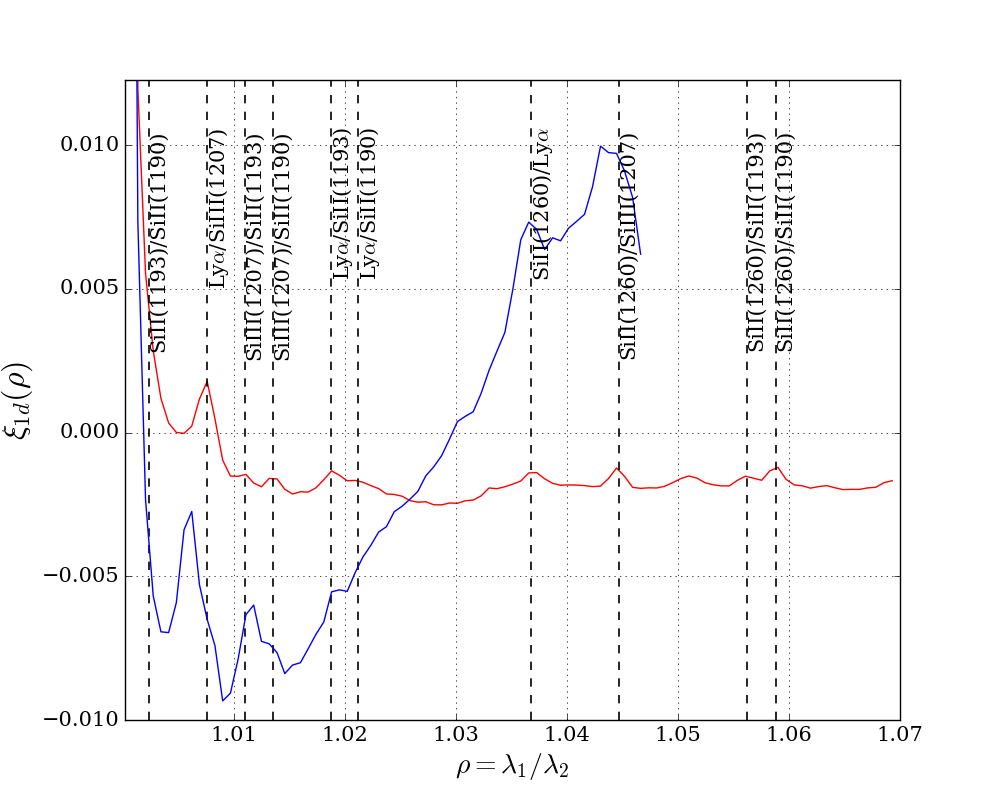}
	\caption{\label{fig:c1d_auto} The one-dimensional correlation functions, $\xioned$, in the Ly$\alpha$ (red curve) and Ly$\beta$ (blue curve) regions as a function of the ratio of transition wavelengths. Peaks are due to absorption by the two labeled elements at zero physical separation (Table \ref{table:metal_auto}).}
\label{fig:xi1d}
\end{figure} 

\begin{table}[t]
  \caption{The Ly$\alpha$/metal and metal/metal pairs contributing
    to the flux correlation function.
    The table shows the ratio of transition wavelengths
    and the corresponding apparent separation, $\rpar^{\rm ap}$,
for  pairs at vanishing
physical separation, computed at an average redshift of 2.34
using eqn. (\ref{eq:rapparent})
  }
  \label{table:metal_auto}       
  \centering
  \begin{tabular}{ccc}        
    \hline
    \hline
    Transitions & $\lambda_1/\lambda_2$ & $\rpar^{\rm ap}$[\hMpc]\\
    \hline
    \ion{Si}{II}(1193)/\ion{Si}{II}(1190) & 1.002 & 7 \\
    Ly$\alpha$(1216)/\ion{Si}{III}(1207) & 1.008 & 21 \\
    \ion{Si}{III}(1207)/\ion{Si}{II}(1193) & 1.011 & 31 \\
    \ion{Si}{III}(1207)/\ion{Si}{II}(1190) & 1.014 & 38 \\
    Ly$\alpha$(1216)/\ion{Si}{II}(1193) & 1.019 & 52 \\
    Ly$\alpha$(1216)/\ion{Si}{II}(1190) & 1.021 & 59 \\
    \ion{Si}{II}(1260)/Ly$\alpha$(1216) & 1.037 & 105 \\
    \ion{Si}{II}(1260)/\ion{Si}{III}(1207) & 1.045 & 126 \\
    \ion{Si}{II}(1260)/\ion{Si}{II}(1193) & 1.056 & 157 \\
    \ion{Si}{II}(1260)/\ion{Si}{II}(1190) & 1.059 & 164 \\
        \hline        
        \end{tabular}
\end{table}

\subsection{The flux-transmission field $\delta_q(\lambda)$}
\label{sec:deltas}

The computation of the correlation function requires an estimation of the
transmission field along the line-of-sight (LOS) towards surveyed quasars. This field arises due to the presence along the LOS of intergalactic gas.
More precisely, for the correlation calculation, we only need to know the flux fluctuations around the average transmitted flux spectrum in the forests of quasar $q$ at wavelength $\lambda$. We thus define the field $\delta_q(\lambda)$, for each quasar $q$ under investigation, as:
\begin{equation}
\delta_q (\lambda)\equiv \frac{f_q(\lambda)}{C_q(\lambda)\overline{F}(z)} - 1,
\label{eq:delta}
\end{equation}
where $f_q(\lambda)$ denotes the observed flux of quasar $q$ at observed wavelength $\lambda$, $C_q(\lambda)$ is the continuum flux and $\overline{F}(z)$ the mean transmission at the absorber redshift $z$.

We estimate the quantity $C_q\overline{F}(z)$
from the average of the transmitted flux of all forest spectra in the sample:
\begin{equation}
\overline{f}(\lambda_{\rm RF})=\frac{\sum_q w_q(\lambda_{\rm RF}) f_q(\lambda_{\rm RF})}{\sum_q w_q(\lambda_{\rm RF})}.
\label{eq:ave_flux}
\end{equation}
where $\lambda_{\rm RF}$ is the rest-frame wavelength
and $w_q$ is a weight (see \S~\ref{sec:weights}). 
For each quasar, $\overline{f}(\lambda_{\rm RF})$
is then multiplied  by a linear polynomial function of $\Lambda\equiv \log(\lambda)$ to account for the diversity of quasar luminosity and spectral shape:
\begin{equation}
C_q(\lambda)\overline{F}(z) = \overline{f}(\lambda_{\rm RF})(a_q+b_q \Lambda)\;.
\label{eq:fitcont}
\end{equation}
There are thus two adjustable parameters per quasar, $a_q$ and $b_q$.

Those forests with identified DLAs are given a special treatment.
All pixels where the absorption due to the DLA is higher than 20\% are not used.
The absorption in the wings is corrected  using a 
Voigt profile following
the procedure of \citet{Noterdaeme12}.

 The fitting procedure to determine $(a_q,b_q)$ forces to zero the mean
 and spectral slope of  $\delta_q(\lambda)$ for each quasar, thus introducing spurious correlations in the measured field.
To make it easier to deal with this distortion in the analysis,
we follow B17 by transforming
the measured $\delta_q(\lambda)$ to ${\hat\delta}_q$:
\begin{equation}
{\hat \delta}_q(\lambda_i) = \sum_j \eta_{ij}\delta_q(\lambda_j),
\label{eq:delta_transformation1}
\end{equation}
where
\begin{equation}
\eta_{ij}=\delta^K_{ij}-\frac{w_q(\lambda_j)}{\sum_k w_q(\lambda_k)}-(\Lambda_i-{\overline {\Lambda_q}})\frac{w_q(\lambda_j)(\Lambda_j-{\overline {\Lambda_q}})}{\sum_k w_q(\lambda_k)(\Lambda_k-{\overline {\Lambda_q}})^2},
\label{eq:transform_p}
\end{equation}
where $\delta^K_{ij}$ denotes the Kronecker symbol.
The advantage of this transformation is that it makes the distortion of the true field introduced by the continuum fit procedure explicit, and, as a consequence, simplifies the link between the true correlation function and the measured, distorted one (see \S~\ref{sec:dist_mat}). 

  The statistics of $\hat\delta_q(\lambda)$ within individual
  forests are described (in part) by 
the so-called one-dimensional correlation function,
$\xioned(\lambda_1/\lambda_2)=\langle\hat\delta(\lambda_1)\hat\delta(\lambda_2)\rangle$.
Fig. \ref{fig:xi1d} presents this function for the \lya and \lyb forests.
The peaks  are due to absorption by
different transitions at the same physical position.
Table \ref{table:metal_auto} lists the important observed transition
pairs.
(See also \citet{Pieri14metals}).

\subsection{Pixel weights}
\label{sec:weights}

The pixel weights are proportional to the
inverse of the variance of  $\delta_q(\lambda)$.
Following \citet{Blomqvist18}, the variance 
is modeled as the sum of three terms: 
\begin{equation}
\sigma_q^2 (\lambda)= \eta(\lambda) \sigma^2_{\rm noise} + \sigma^2_{\rm LSS}(\lambda) + \epsilon(\lambda)/\sigma^2_{\rm noise}.
\end{equation}
The noise pixel variance is
$\sigma^2_{\rm noise} = \sigma^2_{\rm pip}/(C_q\overline{F})^2$
where $\sigma^2_{\rm pip}$ is the pipeline estimate of the pixel variance.
The intrinsic, redshift dependent, contribution of the
density fluctuations underlying Ly$\alpha$ regions is $\sigma^2_{\rm LSS}$.
The third  term, $\epsilon(\lambda)/\sigma^{2}_{\rm noise}$,
takes into account differences between the fitted quasar spectrum
and the individual spectrum of quasar $q$ (these differences appear at high signal-to-noise).
The functions $\eta(\lambda)$ and $\epsilon(\lambda)$ correct
for imperfections of  the pipeline estimates and differences
between the average and individual spectra, respectively. 

Following \citet{Busca13}, the weights are corrected to take into account
the expected redshift dependence of the  correlation function amplitude:  
\begin{equation}
\label{eq:corrected_w}
w_q(\lambda) = \frac{(\lambda/\lambda_\alpha)^{\gamma_\alpha-1}}{\sigma_q^2(\lambda)}
\end{equation}
where the Ly$\alpha$ bias redshift-evolution parameter,
$\gamma_{\alpha}=2.9$ \citep{McDonald06}
and $\lambda_\alpha$ is the \lya restframe wavelength.

  In practice,
  one starts with an initial estimate of the weights, allowing
  a first estimate of
  the mean spectrum $\bar{f}(\lambda_{\rm RF})$  (eqn. \ref{eq:ave_flux})
and the quasar parameters $a_q$ and $b_q$ (eqn. \ref{eq:fitcont}).
  The functions $\eta(\lambda)$, $\epsilon(\lambda)$ and
$\sigma_{\rm LSS}(\lambda)$ are  then fit
and the mean spectrum 
is then recalculated with the new weights.
This process is repeated until stable values are obtained after about
five iterations.

\subsection{The correlation function}
\begin{figure}
	\includegraphics[width=\columnwidth]{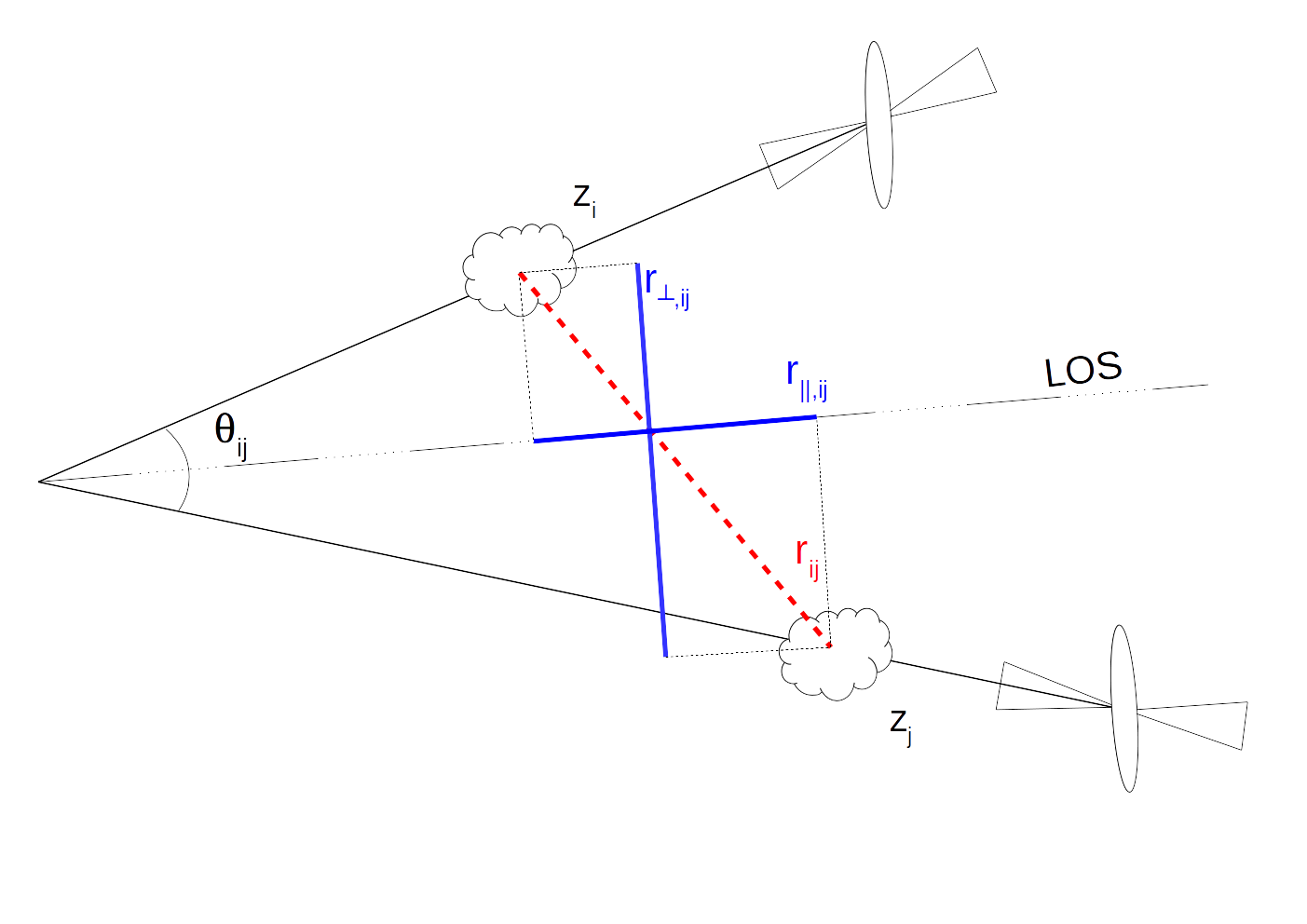}
	\caption{Definition of the coordinates of pixels used in the computation of the correlation function. Absorbers $i$ and $j$ have angular separation $\theta_{ij}$ and distance separation $r_{ij}$. The radial separation $r_{\parallel,ij}$ is the projection of $r_{ij}$\
	  on the median LOS and the transverse separation $r_{\perp,ij}$ is the LOS perpendicular component of $r_{ij}$, assuming
          the flat Pl2015 model (Table \ref{tab:fidcosm}).}
	\label{fig:pairij}
 \end{figure}

To compute the correlation function, we correlate absorption
at an observed wavelength $\lambda_i$ in the LOS of a given quasar $q$,
with   absorption at an observed wavelength $\lambda_j$ in the LOS of another quasar $q'$.
Assuming the absorption is due to the Ly$\alpha$ transition,
one can compute, from the values of $\lambda_i$ and  $\lambda_j$,
the redshifts $z_i$ and $z_j$ of the matter absorbing these lines.
Each pair of absorbers $(z,q)$ entering the computation defines a "pixel"
in real space and we call $r_{ij}$ the physical separation
between two such pixels $i$ and $j$ (see Fig. \ref{fig:pairij}).
This distance is calculated assuming the Pl2015 cosmology
(Table \ref{tab:fidcosm}).
The distance $r_{ij}$ can be projected on the radial and the transverse
directions, leading to two components $r_{\parallel,ij}$ and $r_{\perp,ij}$.
These components can be expressed in terms of the comoving distances $D(z_i)$ and $D(z_j)$ from us to absorbers $i$ and $j$ and the subtended angle between the two LOS, $\theta_{ij}$, as:

\begin{equation}
\left\{
\begin{array}{ll}
r_{\parallel,ij} = \left(D(z_j)-D(z_i)\right)\cos\left(\frac{\theta_{ij}}{2}\right)  \\
r_{\perp,ij} =  \left(D(z_i)+D(z_j)\right)\sin\left(\frac{\theta_{ij}}{2}\right)
\end{array}
\right..
\label{eq:rprt}
\end{equation}

We then define bins of ($r_{\parallel,ij},r_{\perp,ij}$) on a 2D grid.
In practice, the grid uses 2500 bins of dimensions
$4h^{-1}{\rm Mpc}\times 4h^{-1}{\rm Mpc}$
over $0<\rperp<200$\hMpc~and $0<\rpar<200$\hMpc.
For a given bin in this grid, $A$, we consider each pair of pixels ($i$,$j$) whose $r_{\parallel}$ and $r_{\perp}$ coordinates fall on this bin. The measured correlation function in bin $A$ reads:
\begin{equation}
\label{eq:cf}
{\hat \xi}(A) = \frac{\sum_{(i,j)\in A} w_i w_j {\hat \delta}_i {\hat \delta}_j}{\sum_{(i,j)\in A} w_i w_j},
\end{equation}	
with $w_k \equiv w_{q_k}(\lambda_k)$ and ${\hat \delta}_k\equiv {\hat \delta}_{q_k}(\lambda_k)$.

We discard from the computation all pixel pairs belonging to the same LOS, since two pixels belonging to the same quasar spectrum are affected in a correlated way by the fitting procedure described in \S \ref{sec:deltas}. Likewise, pixels belonging to the same half plate at the same wavelength are excluded, to avoid unphysical correlations induced by the extraction pipeline.

\subsection{The covariance matrix}
\label{subsec:cov}

\begin{figure*}
	\includegraphics[width=.5\textwidth]{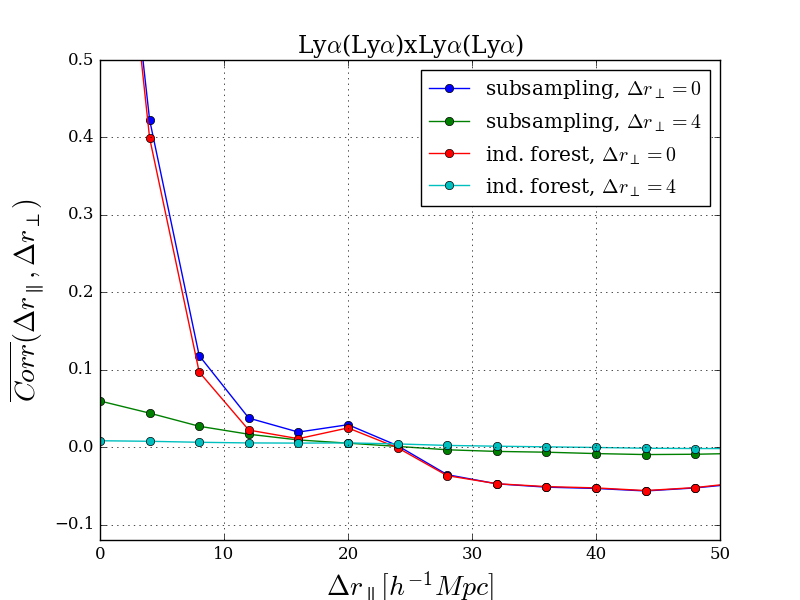}
	\includegraphics[width=.5\textwidth]{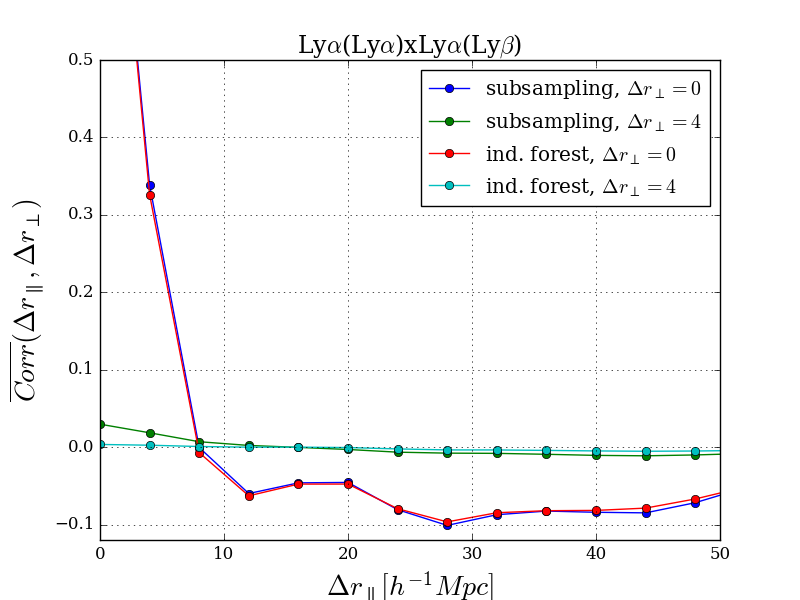}
	\caption{\label{fig:corr_matrices}
          Averaged correlations, $Corr_{AB} = C_{AB} / \sqrt{C_{AA} C_{BB}}$, vs. $\Delta r_{\parallel}$ for the two lowest intervals of $\Delta r_{\perp}$,
          for the \autolya (left) and \crosslya (right) correlation functions.
The subsampling covariances are calculated using (\ref{eq:covmat_estimate})
and the independent-forest estimates by (\ref{eq:xi1dcov}).        }
\end{figure*} 

\begin{figure}
        \includegraphics[width=\columnwidth]{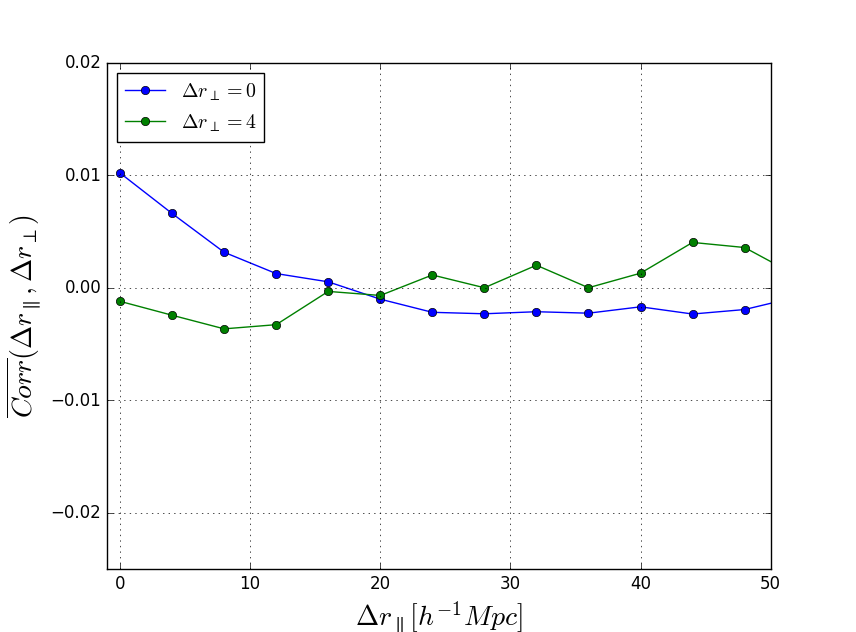}
	\caption{\label{fig:corr_matrices_cross}
          Averaged correlations, $Corr_{AB} = C_{AB} / \sqrt{C_{AA} C_{BB}}$, vs. $\Delta r_{\parallel}$ for the two lowest intervals of $\Delta r_{\perp}$,
          for the cross-covariance matrix between \autolya and \crosslya correlation functions.
        }
\end{figure} 

The covariance between two bins $A$ and $B$ is defined as: 
\begin{equation}
  C_{AB} = \Big\langle {\hat \xi}_A {\hat \xi}_B\Big\rangle -
  \Big\langle {\hat \xi}_A \Big\rangle \Big\langle {\hat \xi}_B \Big\rangle,
\label{eq:covmat}
\end{equation}
where $\langle....\rangle$ denotes an ensemble average.
Following \citet{Delubac15} and B17, we estimate equation (\ref{eq:covmat}) by dividing the eBOSS footprint in $N_h=876$ sky pixels, using the HEALPix tessellation scheme (see \citealt{Gorsky05}),
and by equating the ensemble averages of equation (\ref{eq:covmat})
with  the weighted mean over these sky pixels: 
\begin{equation}
\label{eq:mean}
\Big\langle {\hat \xi}_A \Big\rangle \approx \frac{\sum_h W_A^h {\hat \xi}_A^h}{\sum_h W_A^h},
\end{equation}
and 
\begin{equation}
\label{eq:mean2}
\Big\langle {\hat \xi}_A {\hat \xi}_B \Big\rangle \approx  \frac{\sum_{h} W_A^h W_B^{h} {\hat \xi}_A^h {\hat \xi}_B^{h}}{\Big (\sum_{h} W_A^h \Big ) \Big (\sum_{h} W_B^{h} \Big)} = W_A^{-1} W_B^{-1} \sum_{h} W_A^h W_B^{h} {\hat \xi}_A^h {\hat \xi}_B^{h},
\end{equation}
with $W_A^h$ the sum of the weights of pairs in sky pixels $h$ contributing to bin $A$. Similarly, $\xi^h_A$ is the correlation function of pairs in sky pixels $h$ that contribute to bin $A$.

In practice, for the computation of the correlation function, a pair $(i,j)$ is attributed to the sky pixel of the first quasar of the pair, and the pair $(j,i)$ is never considered, insuring that a pair is not counted twice in the calculation.

In this approximation, we assume that each sky pixel provides an independent realization of the $\delta$ field.
This statement is not exactly true as correlations do exist between pairs in different
sky pixels, but these correlations are small (e.g., \citealt{Delubac15}).

We thus compute the covariance matrix defined in equation (\ref{eq:covmat}) using the following expression:
\begin{equation}
\label{eq:covmat_estimate}
C_{AB} =   \frac{\sum_h W_A^h W_B^h ({\hat \xi^h}_A {\hat \xi}_B^h - {\hat \xi}_A {\hat \xi}_B)}{\Big (\sum_{h} W_A^h \Big ) \Big (\sum_{h} W_B^{h} \Big)},
\end{equation}
where $\hat{\xi}_A$ is given by (\ref{eq:cf}).
Due to the finite number of sky pixels,
the estimate (\ref{eq:covmat_estimate}) is noisy and must be smoothed
before it can be used in fits. 
We perform the smoothing  by approximating the correlation,
$Corr_{AB} = C_{AB}/\sqrt{C_{AA}C_{BB}}$,
 as a function of $\Delta\rpar = |\rpar^A - \rpar^B|$ and
  $\Delta\rperp = |\rperp^A - \rperp^B|$ only,
  ignoring the small dependence on $\rpar$ and $\rperp$. 

As a check of the subsampling method,
the covariance
can also be estimated by 
neglecting inter-forest correlations, 
in which case the four-point function vanishes unless
the four pixels are drawn from just two spectra:
\begin{equation}
C_{AB} =
\frac{1}{W_AW_B}
\sum_{ij\in A}\sum_{kl\in B} w_iw_j w_kw_l 
\xioned(\lambda_i/\lambda_k)\xioned(\lambda_j/\lambda_l)
\label{eq:xi1dcov}
\end{equation}
where $\xioned$ is the intra-forest correlation function shown in
Fig. \ref{fig:xi1d}.
The sum can then be estimated
from a random sample of forest pairs.
  Because neighboring forests are nearly parallel, the sum
  necessarily produces $C_{AB}=0$ unless $\rperp^A\sim\rperp^B$.

  Because the \lya and \lyb forests have different $\xioned$,
we expect differences between the covariances for
\autolya and \crosslya correlations.
These differences are illustrated  in Fig. \ref{fig:corr_matrices}
showing, for the two
lowest values of $\Delta\rperp$,
the correlation, 
$Corr_{AB}$.
For $\Delta\rperp=0$, there is good agreement between the subsampling
(\ref{eq:covmat_estimate})
and independent-forest (\ref{eq:xi1dcov}) calculations.

  Figure \ref{fig:corr_matrices_cross} displays the $Corr_{AB}$
  between the \autolya and \crosslya correlation functions.
  The $Corr_{AB}$ are less than a percent, and will be ignored
  in the fits of the correlation functions.

\begin{table}[tb]
\centering
\caption{Parameters of the model of the correlation function.
  The standard-fit parameters are given in the first section
  of the table. The second section lists parameters that are fixed in
  the standard fit, together with their values.
} 
\label{table:lyaparams}
\begin{tabular}{l l}
\hline
\hline
\noalign{\smallskip}
Parameter & Description\\
\noalign{\smallskip}
\hline
\noalign{\smallskip}
$\alpha_{\parallel}, \alpha_{\perp}$ & BAO peak-position parameters\\
$\blya,\betalya$ & Bias parameters for Ly$\alpha$ absorption\\
$\bhcd,\betahcd$ & Bias parameters of HCD systems\\
$b_{\rm m}$ & Bias of metal species\\
\noalign{\smallskip}
\hline
\noalign{\smallskip}
$L_{\rm HCD}$ & Smoothing scale of HCD systems\\
$\;\;\;=10$~\hMpc & \\
$\Sigma_{\perp}=3.26$~\hMpc & Transverse broadening of BAO peak\\
$\Sigma_{\parallel}=6.41$~\hMpc & Radial broadening of BAO peak\\
$\beta_{\rm m}=0.5$ & Redshift-space distortion for  \\
 & \ion{Si}{II}(1190), (1193), (1260), \ion{Si}{III} (1207)\\
$\beta_{\rm{CIV(eff)}}=0.27$ & \ion{C}{IV}(eff) redshift-space distortion \\
$R_{\parallel},R_\perp$ & Binning smoothing parameter\\
$\;\;=4$~\hMpc & \\
$A_{\rm peak}=1$ & BAO peak amplitude\\
$\gamma_{\alpha}=2.9$ & Ly$\alpha$  bias evolution exponent\\
$\gamma_{\rm m}=1$ & Metal  bias evolution exponent\\
\noalign{\smallskip}
\hline
\end{tabular}
\end{table}

\section{Modeling the  correlation function}
\label{sec:model_lya}

This section details the model of the Ly$\alpha$ auto-correlation function that will be fitted against the estimator $\hat\xi_A $ of equation~(\ref{eq:cf}).
Table \ref{table:lyaparams}
lists the different parameters of the model.
The model includes two components: one is the Ly$\alpha$-only 
correlation function computed from 
Ly$\alpha$ absorption only; the other component incorporates the contribution to the correlation function of absorption by metals, for which the nominal separation for the pixel is not the true separation
(see \S \ref{subsec:conta_metal}). We thus write:

\begin{equation}
\label{eq:xi_mod}
\xi_{\rm mod} = \xi_{\rm mod}^{\rm Ly\alpha-Ly\alpha} + \xi_{\rm mod}^{\rm metals}.
\end{equation}

The following subsections describe these two components.
Section \ref{sec:dist_mat} explains how the model is ``distorted''
to fit the data.

\subsection{The baseline model for $\xi_{\rm mod}^{\rm Ly\alpha-Ly\alpha}$}
\label{subsec:baseline}

We start from the CAMB linear power spectrum \citep{Lewis00} which is decomposed into a smooth and a peak components, following the side-band technique described by \citet{Kirkby13}.
This allows one to constrain the position of the BAO peak
independently of the correlation function at scales much smaller or much larger than the BAO distance scale. We thus model the matter power spectrum as the sum of two terms corresponding to the smooth and the peak terms in the correlation function. Moreover, in order to incorporate the effects of the
non-linear growth of matter
that lead to  broadening of the BAO peak,
the peak term is corrected by a Gaussian factor \citep{Eisenstein07}. The ``quasi-linear'' power spectrum hence reads:
\begin{equation}
  P_{\rm QL}(\vec{k},z) = P_{\rm smooth}(\vec{k},z) +
  \exp
  \left(-\frac{k_{\parallel}^2 \Sigma_{\parallel}^2}{2}
        -\frac{k_{\perp}^2 \Sigma_{\perp}^2}{2} \right)
  P_{\rm peak}(\vec{k},z),
\label{eq:pql}
\end{equation}
where $P_{\rm smooth}(\vec{k},z)$ and $P_{\rm peak}(\vec{k},z)$
are the power spectra of the smooth and peak components, and $\Sigma_{\parallel}$ and $\Sigma_{\perp}$ represent the RMS displacements in the parallel and transverse directions, respectively. We adopt the values of \citet{Kirkby13} for these parameters:  $\Sigma_{\parallel}= 6.41$\hMpc~ and $\Sigma_{\perp}= 3.26$\hMpc. 

The power spectrum is obtained from $P_{\rm QL}$ as
\begin{equation}
\label{eq:Pk_alpha}
P_{\rm Ly\alpha-Ly\alpha}(\vec{k},z) = P_{\rm QL}(\vec{k},z) \dlya^2(\vec{k},z) D_{\rm NL}(\vec{k}) G(\vec{k}),
\end{equation}
where $\dlya$ is the Kaiser factor \citep{Kaiser87} for the Ly$\alpha$ absorption and
$D_{\rm NL}(\vec{k})$ takes into account non-linear effects.
The function $G({\bf k})$ models the effect of binning of the
correlation function on the separation grid.

The Kaiser factor can be written as:
\begin{equation}
\label{eq:kais}
\dlya = \blya^\prime (z)(1+\betalya^\prime \mu_k^2),
\end{equation}
where $\blya^\prime$ is the effective bias of Ly$\alpha$
absorbers with respect to the underlying matter density field,
$\betalya^\prime $ is
the effective redshift space distortion (RSD) parameter, 
and $\mu_k=k_\parallel/k$.
The two effective parameters ($\blya^\prime$ and $\blya^\prime\betalya^\prime$)
combine  \lya absorption in the IGM and
in unmasked high-column density
(HCD) systems, i.e., HI absorbers with column densities
$N_{HI}>10^{17.2}\mathrm{cm}^{-2}$:
\begin{equation}
\left\{
\begin{array}{ll}
\blya^\prime = \blya  + \bhcd \Fhcd (\kpar) \\[2mm]
\blya^\prime\betalya^\prime =
\blya \betalya  + \bhcd \betahcd\Fhcd (\kpar)~~\raisebox{8pt}{,}
\end{array}
\right.
\end{equation}
where $(\blya,\betalya)$ and
$(\bhcd,\betahcd)$ are the bias parameters associated with
the IGM and HCD systems and $\Fhcd$ is a function defined below.

Following \citet{McDonald06}, we assume that the product of
$\blya $ and the growth factor of structures varies with redshift as
$(1+z)^{\gamma_{\alpha}-1}$, with $\gamma_{\alpha} = 2.9$,
while  we make use of the approximation that
$\betalya $ does not depend on redshift.

HCD absorbers
are  expected to trace the 
underlying density field and
their effect on the flux-transmission field depends on whether 
they are identified and given the special treatment
described in Sect. \ref{sec:data}.
If they are correctly identified with the total absorption region masked
and the wings correctly modeled, they can be expected to have no
significant effect on the field.
Conversely, if they are not identified, the measured
correlation function will be modified because
their absorption is spread  along the radial direction.
This broadening effect
introduces a $\kpar$ dependence of the effective bias
\citep{Font-Ribera12}.
Following the study of  \citet{Rogers18},
we adopt a simple exponential form, $\Fhcd =\exp(-\Lhcd \kpar)$, 
where $\Lhcd $ is a typical length scale for these systems.
DLA identification is possible if their width
(wavelength interval for absorption greater than 20\% )
is above $\sim2.0$ nm, corresponding to $\sim14$\hMpc~in our sample.
Based on results from \citet{Rogers18} results, we impose $\Lhcd \sim 10$\hMpc~while fitting for the
bias parameters $\bhcd $ and $\betahcd$.  Fixing $\Lhcd$ is necessary because otherwise the model becomes too unconstrained.
We have verified that setting $\Lhcd$ in the range $7 <\Lhcd< 13$~\hMpc~
does not significantly change the inferred BAO peak position.

We focus on the minimal model able to reproduce the data, designated as "baseline model". This baseline model does not include the correction of the
UV background fluctuations \citep{Pontzen14,GontchoAGontcho14}
used in B17. We discuss the improvement of the fit when this UV correction is added, in section \ref{sec:fit}.

The function $D_{\rm NL}(\vec{k})$
accounts for non-linear effects such as thermal broadening,
peculiar velocities and non-linear structure growth. A fitting formula for $D_{\rm NL}$ is given by equation (21) of \citet{McDonald03} and has been extensively used in previous studies. More recently, \citet{Arinyo15} proposed a new fitting formula involving  6 free parameters given by
their equation (3.6).
Besides reducing number of free parameters with respect
to \citet{McDonald03}, it has the correct behavior at small wavenumber $k$ and an explicit dependence on $P_{\rm QL}(k)$, whereas this dependence is only implicit
in the \citet{McDonald03} formula.
In practice, the two approaches yield similar results but
for the above reasons, we adopt  the formula of \citet{Arinyo15} in the present work and linearly interpolate the parameter values from their Table 7 at the effective redshift $z=2.34$.

To account for the effect of the binning of the correlation function on the separation grid, we assume the distribution to be homogeneous on each bin\footnote{In fact, in the perpendicular direction the distribution is approximately proportional to $r_\perp$; however, assuming homogeneity produces a sufficiently accurate correlation function (B17).} and compute the function $G(\textbf{k})$ as 
the product of the Fourier transforms of the rectangle functions that model a uniform square bin: 
\begin{equation}
G(\textbf{k}) = \sinc(\frac{k_{\parallel} R_{\parallel}}{2})  \sinc(\frac{k_{\perp} R_{\perp}}{2}),
\end{equation}
where $R_{\parallel}$ and $R_{\perp}$ are the radial and transverse widths of the bins, respectively. \\

The two terms in $P_{\rm QL}(\vec{k},z)$  (eqn. \ref{eq:pql}) are
Fourier transformed to the smooth and peak components
of the correlation function:
\begin{equation}
\label{eq:xi_mod_lyalya}
\xi_{\rm mod}^{\rm Ly\alpha-Ly\alpha}(r_{\parallel},r_{\perp},\alpha_{\parallel},\alpha_{\perp}) = \xi_{\rm smooth}(r_{\parallel},r_{\perp}) + A_{\rm{peak}}~\xi_{\rm peak}(\alpha_{\parallel}r_{\parallel},r_{\perp}\alpha_{\perp}).
\end{equation}
The amplitude of the peak, $A_{\rm{peak}}$, is fixed to unity in the standard fit.
In the peak component we have introduced the parameters $(\apar,\aperp)$ which
allow us to fit for the peak position independently of the smooth component:
\begin{equation}
\label{eq:aperp_atan}
\alpha_\parallel = \frac{D_H(\overline{z})/r_d}{[D_H(\overline{z})/r_d]_{\mathrm{fid}}},\\
\alpha_\perp = \frac{D_M(\bar{z})/r_d}{[D_M(\bar{z})/r_d]_{\mathrm{fid}}},
\end{equation}
where $\overline{z}$ is the effective redshift of the measurement and the
suffix 'fid' denotes the Pl2015 cosmology from Table \ref{tab:fidcosm}.

\subsection{The contamination by metals}
\label{subsec:conta_metal}

The second term in the model correlation function (eq. \ref{eq:xi_mod})
accounts for absorption by metals along the quasar LOS.
Such absorption is correlated with \lya absorption \citep{Pieri14metals}
and can be used as a tracer of the density field
\citep{Blomqvist18,Helion19}.
Here, it is a complicating factor in the analysis because
the redshifts of pixels are calculated assuming \lya absorption.

The important metals can be seen in the  1D
correlation function, $\xioned(\lambda_1/\lambda_2)$, shown in 
Fig. \ref{fig:c1d_auto}.
Column 2 of Table \ref{table:metal_auto} lists the wavelength ratios for the main metal/metal and metal/\lya absorption correlations, relevant for the  \lya auto-correlation function computation. 
The corresponding apparent radial separation at
vanishing physical separation is
\begin{equation}
  \rpar^{\rm ap} \approx (1+\overline{z})D_H(\overline{z})
  \frac{\lambda_1-\lambda_2}{\lambda_\alpha}
  \label{eq:rapparent}
\end{equation}
where $\overline{z}$ is the mean redshift of the pair.
Values are  given in Table \ref{table:metal_auto} for $\overline{z} = 2.34$.

We model the power spectrum of each pair of absorbers, $(m,n)$,
with the same form as that for \lya-\lya absorption  (\ref{eq:Pk_alpha})
except that HCD effects are neglected:
\begin{equation}
  \label{eq:PS_mn}
  P_{mn}(\vec{k},z) = b_m b_n (1+\beta_m \mu_k^2) (1+ \beta_n \mu_k^2) G(\vec{k}) P_L(\vec{k},z).
\end{equation}
  Since the $b_m$ and $\beta_m$
  are mostly determined near $(\rperp,\rpar) \sim (0,\rpar^{\rm ap})$,
  they cannot be determined separately.
  We therefore
  fix $\beta_{\rm CIV(eff)} = 0.27$ \citep{Blomqvist18}.
  For the other metal species we keep the value $\beta = 0.50$
  used in \citet{Bautista17}  which comes from DLA
  measurements \citep{Font-Ribera12}.

  The Fourier transform of $P_{mn}(\vec{k},z)$ is then
the model correlation function of the  pair
$(m,n)$: $\xi_{\rm mod}^{m-n}(\tilde{r}_\parallel,\tilde{r}_\perp)$,
where $(\tilde{r}_\parallel,\tilde{r}_\perp)$ are the separations calculated using
the correct restframe wavelengths, $(\lambda_m,\lambda_n)$.

Since we assign a redshift, $z_\alpha$, assuming Ly$\alpha$ absorption, the rest-frame wavelength we ascribe to a metal transition $m$ observed at wavelength $\lambda_i$ is not equal to the true rest-frame wavelength $\lambda_i/(1+z_m)$, where $z_m$ is the true redshift of the metal absorber. This misidentification results in a shift of the model contaminant correlation function. For each pair $(m,n)$ of contaminants, we compute the shifted model correlation function
with respect to the unshifted
model metal correlation function $\xi_{\rm mod}^{m-n}$
by introducing a metal matrix  $M_{AB}$ \citep{Blomqvist18}, such that:
\begin{equation}
  \xi_{\rm mod}^{m-n}(A) = \sum_B M_{AB}
  \xi_{\rm mod}^{m-n}(\tilde{r}_\parallel(B),\tilde{r}_\perp(B)),
\end{equation}
where:
\begin{equation}
M_{AB}=\frac{1}{W_A}\sum_{(m,n)\in A,(m,n)\in B}w_mw_n,
\end{equation}
and $(m,n)\in A$ refers to pixel separation computed assuming $z_\alpha$,
and $(m,n)\in B$ to pixel separation computed using
the redshifts of the $m$ and $n$ absorbers, $z_m$ and $z_n$.
We take into account the redshift dependence of the weights, equation (\ref{eq:corrected_w}),  in the computation of $w_m$ and $w_n$.

The total metal contaminant correlation function, $\xi_{\rm mod}^{\rm metals}$, is the sum of all the $\xi_{\rm mod}^{m-n}$ contributions, where $(m,n)$ runs over all the involved transition pairs for the Ly$\alpha$ auto-correlation function, see Table \ref{table:metal_auto}:

\begin{equation}
  \label{eq:metal_cf}
  \xi_{\rm mod}^{\rm metals}(A) = \sum_{m,n} \xi_{\rm mod}^{m-n}(A).
\end{equation}

\begin{table*}
  \centering
  \caption{\label{table:standard_results}
    The parameters of the model of the correlation
    function and the best fit values of the
    \autolya data (third column)
    and to the \autolya and \crosslya data (fourth column). Errors on parameters correspond to $\Delta \chi^2 = 1$.
  }
  \begin{tabular}{lccc}
    \hline
    \hline
    Parameter description &  & \autolya & \autolya\\
        &        &  & +\crosslya \\
    \hline 
 &    $n_{pairs}$ &$5.44 \times 10^{11}$&$6.94 \times 10^{11}$\\
 &   $\sum w_{pairs}$ & $3.56 \times 10^{13}$&$4.20 \times 10^{13}$\\ 
    \hline			
Radial BAO peak-position &    $\alpha_{\parallel}$ & 1.047 $\pm$ 0.035 & 1.033 $\pm$ 0.031  \\
Transverse BAO peak-position&    $\alpha_{\perp}$ & 0.969 $\pm$ 0.041 & 0.953 $\pm$ 0.042  \\
    \hline
\lya redshift-space distortion &    $\betalya$ & 1.773 $\pm$ 0.066 & 1.933 $\pm$ 0.100  \\
 \lya velocity bias $b_{\eta Ly\alpha} = b_{ Ly\alpha}f/\betalya$&    $b_{\eta Ly\alpha}$&-0.208 $\pm$ 0.004 & -0.211 $\pm$ 0.004\\
    \hline
HCD redshift-space distortion &    $\betahcd$ & 0.845 $\pm$ 0.157 & 1.031 $\pm$ 0.153  \\
HCD bias \autolya&    $\bhcd^{ \rm Ly\alpha(Ly\alpha)\times Ly\alpha(Ly\alpha)}$ & -0.047 $\pm$ 0.003 & -0.051 $\pm$ 0.004 \\
HCD bias \crosslya & $\bhcd^{ \rm Ly\alpha(Ly\alpha) \times Ly\alpha( Ly\beta)}$ & -  & -0.072 $\pm$ 0.005  \\
    \hline
Metal absorption bias &    $b_{SiII(1190)}$ & -0.0051 $\pm$  0.0010 &-0.0050 $\pm$  0.0010 \\
&    $b_{\rm SiII(1193)}$ &-0.0046 $\pm$  0.0010 & -0.0046 $\pm$  0.0010\\
&    $b_{\rm SiIII(1207)}$ & -0.0082 $\pm$  0.0010& -0.0080 $\pm$  0.0010\\
&    $b_{\rm SiII(1260)}$ &-0.0025 $\pm$  0.0013 & -0.0022 $\pm$  0.0013\\
&    $b_{\rm CIV(eff)}$ &-0.0185 $\pm$  0.0078& -0.0163 $\pm$  0.0089\\
    \hline
&    $\chi^2_{min}$ &1619.77 & 3258.92 \\
&    DOF & 1590-11 & 3180-12\\
&    Probability & 0.232 & 0.127\\
&    $\chi^2(\alpha_\parallel=\alpha_\perp=1)$ & 1621.55 & 3260.54\\
    \hline				
  \end{tabular}
  \centering
\label{table:baselineresults}
\end{table*}

\subsection{The distorted model}
\label{sec:dist_mat}

  The model correlation function, $\xi_{\rm mod}$ of eqn. (\ref{eq:xi_mod}),
  cannot be fit directly to the estimated correlation function (\ref{eq:cf})
  because
the measured $\hat\delta(\lambda)$ are only related to the true $\delta(\lambda)$
through the transformation (\ref{eq:delta_transformation1}).
Following B17, we can account for this effect in the fit by using a distorted model:
\begin{equation}
\hat\xi_{\rm mod}(A)=\sum_B D_{AB} \xi_{\rm mod}(B),
\end{equation}
  where $D_{AB}$ is
  the distortion matrix which, following equation (\ref{eq:transform_p}), 
  is given by
\begin{equation}
D_{AB} =  W_A^{-1} \sum_{ij\in A} w_i w_j\; \Big(\sum_{i'j' \in B} \eta_{ii'} \eta_{jj'}\Big).
\end{equation}
The accuracy of this method of accounting for the distortion
of the correlation function was
tested with mock data sets by \citet{Bautista17}.

In practice, to avoid prohibitive computational time, the distortion matrix is computed using only a random $5\%$ portion of the total number of pairs.

\begin{figure*}[t]
\centerline{\includegraphics[width=\textwidth]{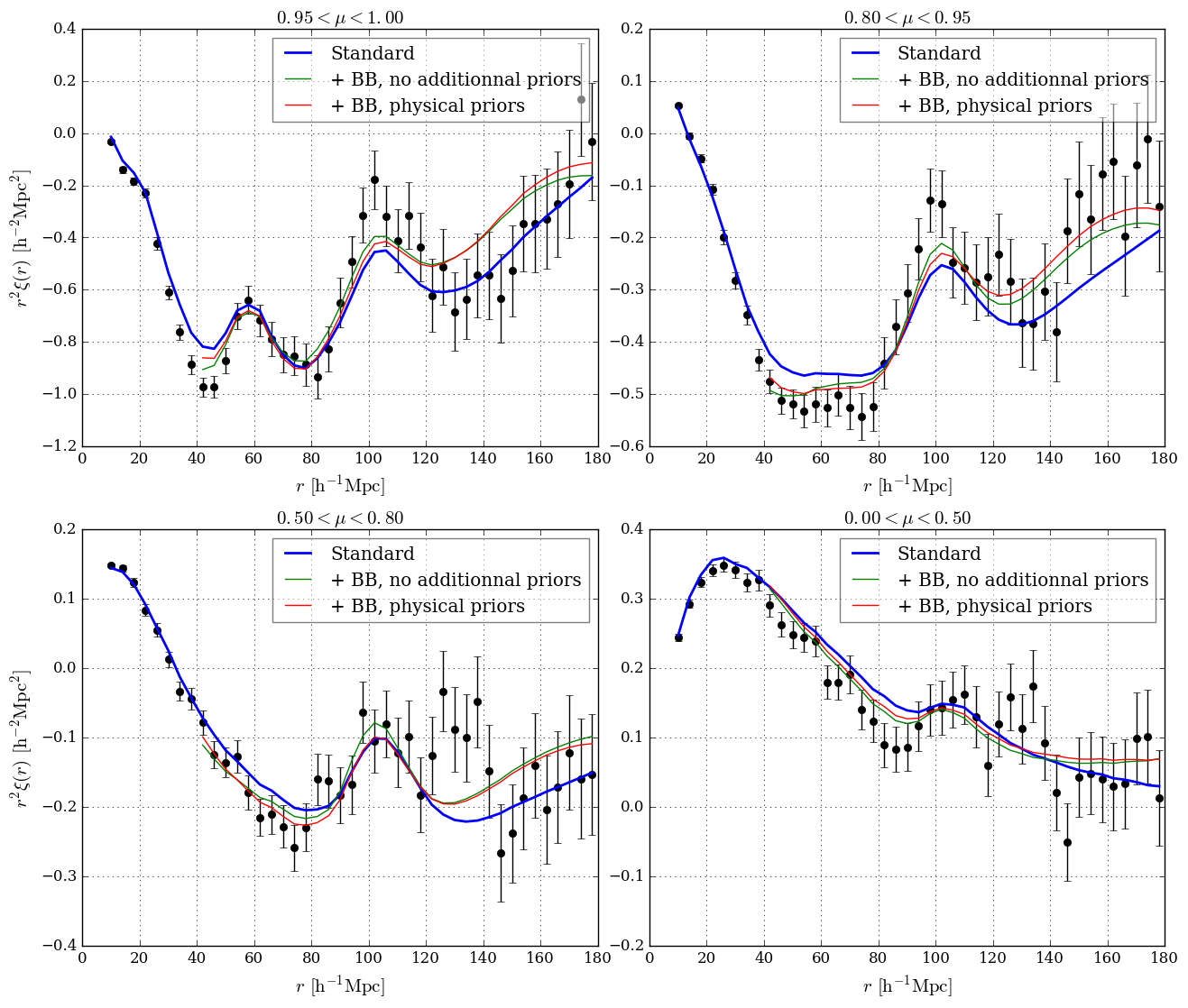}}
\caption{Weighted combination between measured Ly$\alpha$(Ly$\alpha$)xLy$\alpha$(Ly$\alpha$) and Ly$\alpha$(Ly$\alpha$)xLy$\alpha$(Ly$\beta$) correlation functions along with the model best fits in four ranges of $\mu=r_\parallel/r$.
The curves show the standard fit and the two fits with broadband terms
defined by eqn. \ref{xibroadbandeq} with $(\imin,\imax,\jmax)=(0,2,6)$
with and without additional priors, as described in the text.}
\label{figure:aaabwedges}
\end{figure*}

\begin{table*}
  \centering
  \caption{\label{table:model_cf}
    Best fit values of $(\alpha_{\parallel},\alpha_{\perp})$
for the \alllya correlation function fit with various models.
    The first group includes physical models starting with the basic Kaiser redshift-space model and then including, progressively,  metals, HCD, and UV corrections.
    Fits in the second group include polynomial broadband terms,
    as described in the text.
  }
  \begin{tabular}{lcccc}
    \hline
    \hline
    \vspace*{1mm}
    Models &  $\alpha_{\parallel}$ & $\alpha_{\perp}$ & $\chi^2/DOF$ & Probability \\
    \hline
    Kaiser & 1.021 $\pm$  0.028 & 0.977 $\pm$  0.040 & 3624.74/(3180-4) & $3.46 \times 10^{-8}$\\
    \hspace*{3mm}+Metals &1.025 $\pm$ 0.032 & 0.979 $\pm$ 0.044 &  3607.96/(3180-9)&$7.14 \times 10^{-8}$ \\
    \hspace*{3mm}+HCD (baseline) & 1.033 $\pm$ 0.031 & 0.953 $\pm$ 0.042 &  3258.92/(3180-12)& 0.127\\
    \hspace*{3mm}+UV & 1.033 $\pm$ 0.031 & 0.953 $\pm$ 0.042 &  3258.84/(3180-13) & 0.125 \\
    \hline
    BB \\
    Physical priors on $(\blya,\betalya,\bhcd)$&1.037 $\pm$ 0.028 & 0.972 $\pm$ 0.040 & 3006.25/(3030-36) & 0.434\\
    No additional priors & 1.032 $\pm$ 0.027 & 0.980 $\pm$ 0.039 & 3001.00/(3030-36) & 0.460 \\
    \hline					
  \end{tabular}
  \centering
\end{table*}

\section{Fitting the BAO peak position}
\label{sec:fit}

Table \ref{table:baselineresults} presents the best-fit parameters for the
\autolya correlation function alone and
those including the \lyb region, i.e. 
the \alllya correlation function.
Figure \ref{figure:aaabwedges} displays data for the latter 
in four ranges of $\mu$ along with the best fits.
The BAO peak is apparent for $\mu>0.8$ and is suggested for $0.5<\mu<0.8$.

\begin{figure*}
\includegraphics[width=0.5\textwidth]{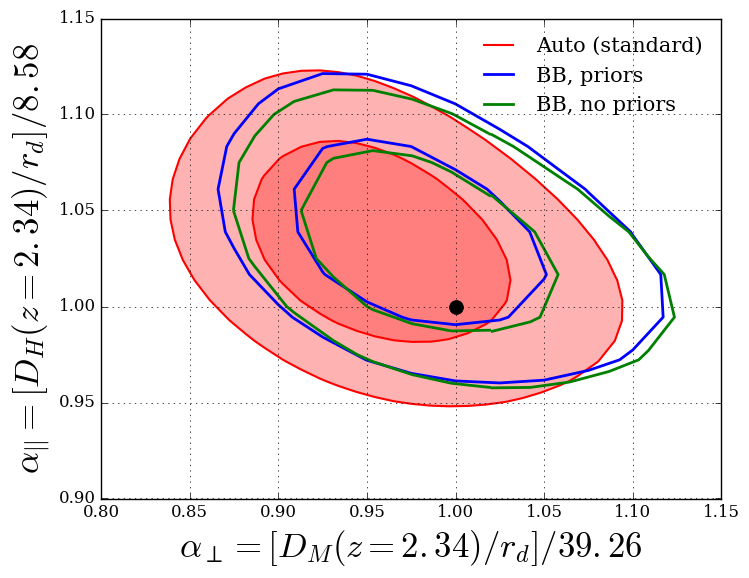}
\includegraphics[width=0.5\textwidth]{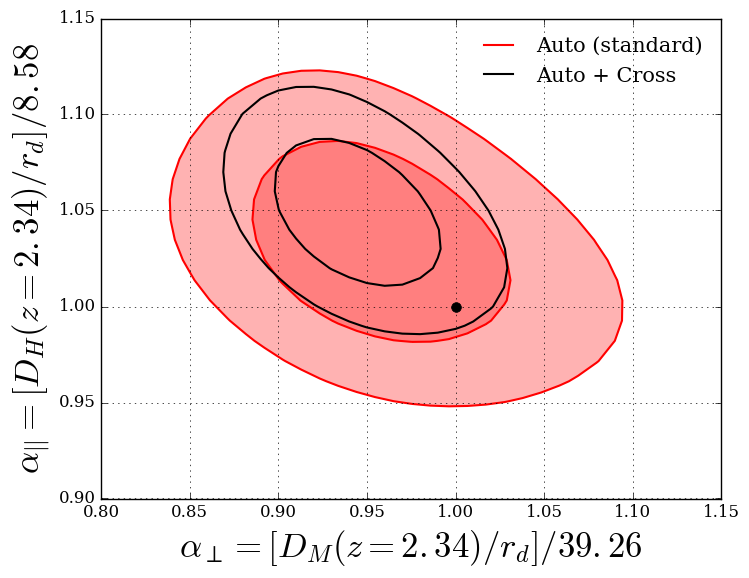}
\caption{\label{fig:contours_auto}
  The left panel shows
  the 68\% and 95\% confidence level contours in the ($\alpha_\parallel$,$\alpha_\perp$) plane from the \alllya auto-correlation
  function for the standard  fit
  and for fits with polynomial broadband (BB) terms with and without additional
  priors, as described in the text.
  The right panel shows the contours for \alllya auto-correlation standard fit
  and those from the combined fit of
  the auto-correlation and 
  the quasar-\lya cross-correlation of \citet{Blomqvist19}.
  In both  panels
  the value  for the Pl2015 model  \citep{Planck16cosmo}
  is shown as a black point.
}
\end{figure*}

The BAO parameters for the  fit using both \lya and \lyb regions are
\begin{equation}
\label{eq:alpha_par_per_masked}
\left\{
\begin{array}{ll}
\alpha_\parallel~=~1.033\;^{+0.034}_{-0.034}\;^{+0.071}_{-0.068}\\[4.pt]
\alpha_\perp~=~0.953\;^{+0.050}_{-0.045}\;^{+0.108}_{-0.091}~\raisebox{8pt}{,}
\end{array}
\right.
\end{equation}
Using the $D/r_d$ values for the Pl2015 cosmology in Table \ref{tab:fidcosm},
these values yield
\begin{equation}
\left\{
\begin{array}{ll}
D_H(2.34)/r_d~=~8.86\;^{+0.29}_{-0.29}\;^{+0.61}_{-0.58}\\[4.pt]
D_M(2.34)/r_d~=~37.41\;^{+1.96}_{-1.77}\;^{+4.24}_{-3.57}~\raisebox{8pt}{.}
\end{array}
\right.
\end{equation}
These results  can be compared with those of B17, who found $\apar=1.053\pm0.036$ and $\aperp=0.965\pm0.055$ at $z=2.33$
using only the \lya region.
These values are very near the present results using only the \lya region:
$\apar=1.047\pm0.035$ and $\aperp=0.960\pm0.041$.
Our use of the \lyb region produces consistent results, given the increase
in the data set; the main improvement is that on the precision on $D_M/r_d$  by $\sim$~25\%.

Constraints on the BAO parameters $(\apar,\aperp)$ are presented in Fig.~\ref{fig:contours_auto}.
Following the method introduced and described in detail in \citet{Bourboux17},
we estimate the relation between $\Delta\chi^{2}=\chi^{2}-\chi_{\rm min}^{2}$ and confidence levels (CLs) for the BAO parameters using a large number of simulated correlation functions generated from the best-fit model and the covariance matrix measured with the data. The results of the study, summarized in Table~\ref{table::confidence}, indicate that the (68.27,95.45\%) confidence levels for $(\apar,\aperp)$ correspond to $\Delta\chi^{2}=(2.74,7.41)$ (instead of the nominal values $\Delta\chi^{2}=(2.3,6.18)$). These levels are shown as the red contours in
Fig.~\ref{fig:contours_auto}
for the \alllya fit.
The best fit is within one standard deviation of
the Pl2015 model.

In addition to the baseline fits, we performed a variety of non-standard
fits to verify that our BAO results are robust and independant of the model.
The results of this exercise are given in Table \ref{table:model_cf}.
The first group of fits starts with the simple Kaiser redshift-space model
and then includes progressively metals, HCD absorption, and UV background
fluctuations.
Including HCD absorption is necessary to obtain a good $\chi^2$
but adding UV fluctuations such as those characterized in
\citet{GontchoAGontcho14}
does not improve the fit,
justifying our choice of ignoring the UV issue in the baseline fit.
In the standard fit, the physical parameters $(L_{\rm HCD}, \Sigma_{\perp}, \Sigma_{\parallel}, \beta_m)$ are fixed (see table \ref{table:lyaparams}) in order to avoid degeneracies with other parameters and non-physical values. 
We have verified that letting them free has no impact on the $\alpha_{\parallel}$ and $\alpha_{\perp}$ parameters. \\

An important test of systematic effects in the position of the BAO peak
is performed by adding polynomial ``broadband'' terms to the correlation
function (before distortion).
We follow the procedure and choice of broadband forms used by B17
and adopt the form 
\begin{equation}
B(r,\mu) =
\sum_{j=0}^{\jmax} \sum_{i=\imin}^{\imax} a_{ij} \frac{L_{j}(\mu)}{r^i}
\hspace*{5mm}\; (j\,\rm{even}) ,
\label{xibroadbandeq}
\end{equation}
where the $L_{j}$ are Legendre polynomials.

We want to ensure that the power-law terms model variations of
the slowly-varying part of the correlation function under the
BAO peak.
We therefore perform these fits only over the restricted range
$40<r<180$\hMpc, avoiding introducing undue influence of the $10<r<40$\hMpc~ range
on the amplitudes of the power laws.
Following B17 we fit with
$(\imin,\imax)=(0,2)$ corresponding to a parabola in 
$r^2\xi_{\rm smooth}$ underneath the BAO peak. We set 
$\jmax=6$, giving four values of $j$  corresponding 
to approximately independent broadbands in each of the four angular ranges in
Fig. \ref{figure:aaabwedges}.

We performed the broadband fits in two ways.  The first placed
``physical priors'' on $(\blya,\betalya,\bhcd)$ in the form of
a Gaussian of mean and width of the fit without broadband terms.
Such priors ensured that the broadband terms were relatively small
perturbations to the physical model.
The second type of fit placed no priors on $(\blya,\betalya,\bhcd)$.

The results of these fits are given in Table \ref{table:model_cf}
and Fig.~\ref{fig:contours_auto}.
We see that the addition of such terms does not change significantly $\apar$
but does shift $\aperp$ by $0.5\sigma$ or $0.7\sigma$ for fits with and
without physical priors.
This effect was already seen in B17 but, with less significance.
Figure \ref{fig:contours_auto} shows that in all cases the BAO
peak position is within one standard deviation of the
prediction of the Pl2015 model.

The fits described above of the \autolya and \crosslya correlation
functions are the primary results of this paper.
We also performed fits with two redshift bins, as described in
Appendix \ref{sec:tomo}.
Each of the two redshifts yielded  values of $(\apar,\aperp)$
that are within $1.2\sigma$ of the Pl2015 model.
(Fig. \ref{fig:contours_binsz}).
We also fit the \crosslyb correlation as described
in Appendix \ref{sec:lyblyb}.
Adding the \lyb absorption data does not add a significant signal
to the BAO peak, but it does allow us to measure the \lyb bias parameters.

\begin{table*}
  \caption{\label{table:combined_auto_cross}
    Best fit results of the \alllya correlation function (second column), of the QSO $\times$ Ly$\alpha$(Ly$\alpha$+Ly$\beta$) correlation function given by \citet{Blomqvist19} (third column) and of the two correlation functions (fourth column).Errors on BAO parameters correspond to CL = 68.27\%, while the other parameters have errors corresponding to $\Delta \chi^2 = 1$. The $\sigma_{ \nu}$, $\Delta r_{\parallel}$, $\xi_0^{TP}$ and $A_{rel1}$ parameters are fit on the QSO $\times$ Ly$\alpha$(Ly$\alpha$+Ly$\beta$) correlation function and fully described in \citet{Blomqvist19}}
  \centering		
  \begin{tabular}{cccc}
    \hline
    \hline
    Parameters & Ly$\alpha$(Ly$\alpha$)xLy$\alpha$(Ly$\alpha$+Ly$\beta$) & QSOxLy$\alpha$(Ly$\alpha$+Ly$\beta$) & Ly$\alpha$(Ly$\alpha$)xLy$\alpha$(Ly$\alpha$ +Ly$\beta$) \\
               & & & + QSOxLy$\alpha$(Ly$\alpha$+Ly$\beta$) \\
    \hline
    $\alpha_{\parallel}$ & 1.033 $\pm$ 0.034 & 1.076 $\pm$ 0.042 & 1.049 $\pm$ 0.026 \\
    $\alpha_{\perp}$ & 0.953 $\pm$ 0.048 & 0.923 $\pm$ 0.046 & 0.942 $\pm$ 0.031 \\
    \hline
    $\betalya$ & 1.933 $\pm$ 0.101 & 2.28 $\pm$ 0.31 & 1.994 $\pm$ 0.099 \\
    $b_{\eta Ly\alpha}$& -0.211 $\pm$  0.004 & -0.267 $\pm$ 0.014 & -0.214 $\pm$ 0.004 \\
    $\beta_{\rm QSO}$ & - & 0.257 & 0.209 $\pm$ 0.006 \\			
    \hline
    $\betahcd$ & 1.031 $\pm$ 0.153 & 0.500 $\pm$ 0.200 &  0.972 $\pm$ 0.150 \\
    $\bhcd^{\rm Ly \alpha(Ly \alpha) \times Ly \alpha(Ly \alpha)}$ & -0.051 $\pm$ 0.004 & - &-0.052 $\pm$ 0.004 \\
    $\bhcd^{\rm Ly \alpha(Ly \alpha) \times Ly \alpha(Ly \beta)}$ & -0.072 $\pm$ 0.005 & - & -0.073 $\pm$ 0.005 \\
    $\bhcd^{\rm QSO \times Ly \alpha(Ly \alpha + Ly \beta)}$ & - & -0.000 $\pm$ 0.004 & -0.000 $\pm$ 0.004 \\
    \hline
    $b_{\rm SiII(1190)}$ &  -0.0050 $\pm$  0.0010 &-0.0057 $\pm$  0.0024 &-0.0043 $\pm$  0.0009\\
    $b_{\rm SiII(1193)}$ &  -0.0046 $\pm$  0.0010 &-0.0015 $\pm$  0.0024 &-0.0034 $\pm$  0.0009\\
    $b_{\rm SiIII(1207)}$ &  -0.0080 $\pm$  0.0010 &-0.0117 $\pm$  0.0024 &-0.0083 $\pm$  0.0009\\
    $b_{\rm SiII(1260)}$ &  -0.0022 $\pm$  0.0013 &-0.0022 $\pm$  0.0017 &-0.0019 $\pm$  0.0009\\
    $b_{\rm CIV(eff)}$ &  -0.0163 $\pm$  0.0089 &- &-0.0167 $\pm$  0.0090\\
    \hline
    $ \sigma_{ \nu}$[\hMpc]& - & 7.60 $\pm$ 0.61 & 7.053 $\pm$ 0.357 \\
    $ \Delta r_{\parallel}$[\hMpc] & - & -0.22 $\pm$ 0.32 & -0.169 $\pm$ 0.284\\
    $ \xi_0^{TP}$ & - & 0.276 $\pm$ 0.158 & 0.478 $\pm$ 0.112 \\
    $A_{rel1}$& - & -13.5 $\pm$ 5.8 & -13.573 $\pm$ 4.721 \\
    \hline
    $\chi^2_{min}$ & 3258.91 & 3231.61 & 6499.31 \\
    $DOF$ & 3180-12 & 3180-14 & 6360-18 \\
    Probability & 0.13 & 0.20 & 0.08\\
    $\chi^2(\alpha_{\parallel}=\alpha_{\perp}=1)$ & 3260.54 & 3235.79 & 6504.30\\
    \hline 
  \end{tabular}
  \centering
\end{table*}

\begin{figure}
  \includegraphics[width=\columnwidth]{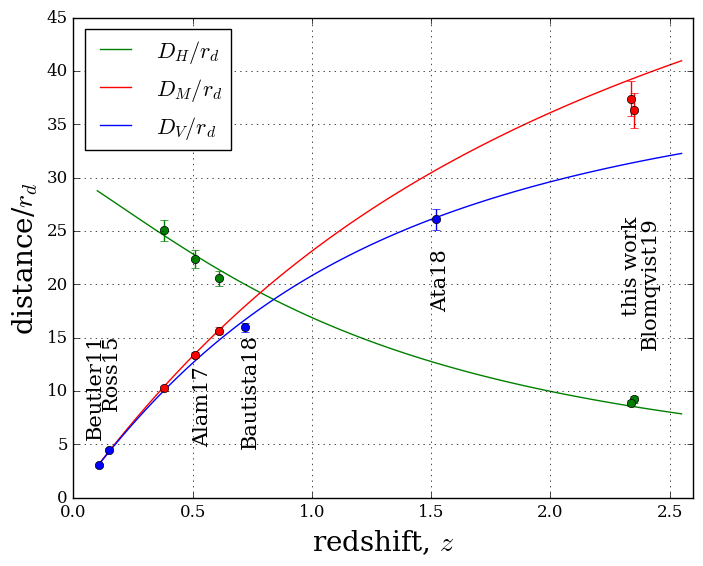}
  \caption{BAO measurement of $D_H/r_d$ and $D_M/r_d$ using BOSS galaxies \citep{Alam17}, Ly$\alpha$ absorption in BOSS-eBOSS quasars (this work) and correlation between BOSS-eBOSS quasars and Ly$\alpha$ absorption \citep{Blomqvist19}. Other measurements give $D_V/r_d$, with $D_V =D_M^{2/3}(zD_H)^{1/3}$, using galaxies (\citet{Beutler11}, \citet{Ross15}, \citet{Bautista18}) and BOSS-eBOSS quasars \citep{Ata18}. Solid lines show the Pl2015 values \citep{Planck16cosmo}.}
  \label{fig:cosmo_dist}
\end{figure}

\begin{figure}
  \includegraphics[width=\columnwidth]{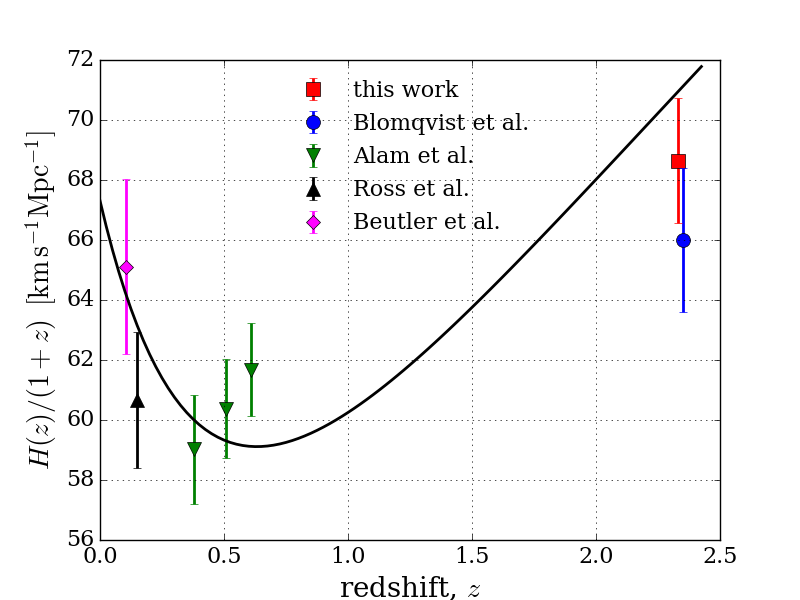}
  \caption{BAO measurement of the comoving expansion rate, $H(z)/(1+z)$, measured with BAO with $r_d = $ 147.3 Mpc. The red square is the present measurement at $z = 2.34$. The measurement by \citet{Blomqvist19} is the blue dot. The other points are computed using galaxy measurements (\citet{Beutler11}, \citet{Ross15}, \citet{Alam17}). The points at $z=0.106$ \citep{Beutler11} and $z=0.15$ \citep{Ross15} are converted from $D_V$ to $H(z)$ using the SNIa measurement of $q_0$ given by \citet{Betoule14}. Solid black line shows the Pl2015 values \citep{Planck16cosmo}.}
  \label{fig:exp_rate}
\end{figure}

Finally,
we combine the measurement of Ly$\alpha$ auto-correlation function of
the present analysis with the Ly$\alpha$ - quasar cross-correlation
measurement of \citet{Blomqvist19} by performing a joint fit of the
two correlation functions. We use the baseline models of the two analyses
and consider the errors to be independent.
The joint fit has 18 free parameters and the effective redshift is $z=2.34$.
The results are given in the column four of
Table \ref{table:combined_auto_cross} and the constraints on $(\aperp,\apar)$
in the right panel of Fig. \ref{fig:contours_auto}.
From this combined fit, we obtain:
\begin{equation}
\left\{
\begin{array}{ll}
\apar~=~1.049\;^{+0.026}_{-0.025}\;^{+0.052}_{-0.051} \\[4.pt]
\aperp~=~0.942\;^{+0.032}_{-0.030}\;^{+0.067}_{-0.059}~\raisebox{8pt}{,}
\end{array}
\right.
\end{equation}
corresponding to:
\begin{equation}
\left\{
\begin{array}{ll}
D_H(2.34)/r_d~=~9.00\;^{+0.22}_{-0.22}\;^{+0.45}_{-0.43}\\[4.pt]
D_M(2.34)/r_d~=~36.98\;^{+1.26}_{-1.18}\;^{+2.63}_{-2.32}~\raisebox{8pt}{.}
\end{array}
\right.
\label{eq:combinedDoverrd}
\end{equation}
The value of $\chi^2$ for $(\apar=1,\aperp=1)$ is 4.99 greater
than the best fit.  Using the confidence levels of
Table \ref{table::confidence}, we conclude that the
results of the combined fit are
1.7$\sigma$ from the predictions of the
Pl2015 model \citep{Planck16cosmo}.

\section{Cosmological constraints}
\label{sec:cosmo}

\begin{figure}[tb]
\centering
\includegraphics[width=\columnwidth]{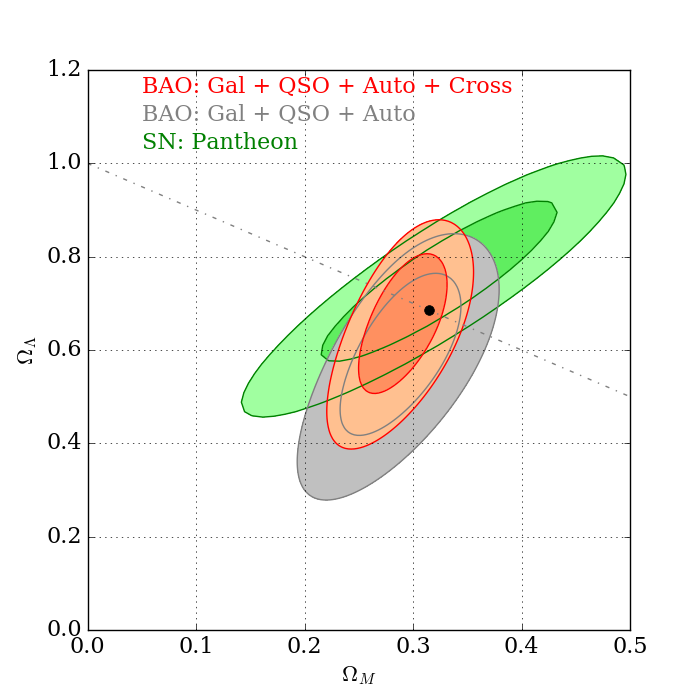}
\caption{
One and two standard deviation constraints on
  $(\Omega_m,\Omega_\Lambda)$.
  The red contours use BAO measurements of  $D_M/r_d$ and $D_H/r_d$
  of this work, of \citet{Blomqvist19} and
  \citet{Alam17},
 and the measurements of $D_V/r_d$ of
  \citet{Beutler11}, \citet{Ross15}, \citet{Ata18} and \citet{Bautista18}.
  The gray contours do not use the \lya-quasar cross-correlation measurement
  of \citet{Blomqvist19}.
  The green contours show the constraints from \Ia~Pantheon sample
  \citep{Scolnic18}. The black point indicates the values for the Planck (2016) best-fit flat $\Lambda$CDM~cosmology.
}
\label{figure::omol}
\end{figure}

BAO data over the redshift range $0.1<z<2.4$ is in overall good agreement
with the predictions of the flat $\Lambda$CDM models consistent with
CMB anisotropies, as illustrated in
Figure \ref{fig:cosmo_dist}.
A striking illustration of the expansion history can be made
by transforming $D_H(z)/r_d$ to $H(z)r_d$.
The measurement presented here gives
\begin{equation}
H(2.34)\frac{r_d}{r_d(Pl2015)} =(227\pm 8)~\mathrm{km}~\mathrm{s}^{-1}~\mathrm{Mpc}^{-1}.
\end{equation}
Fig. \ref{fig:exp_rate} plots this value along with other measurements.
  The data are consistent with the expected behavior of deceleration at high
  redshift followed by acceleration at low redshift.

  Independent of CMB data and without assuming flatness, the BAO data by themselves constrain the parameters $(\Omega_m,\Omega_\Lambda,H_0r_d)$ of the (o)$\Lambda$CDM~model. Using the combined fit (eqn. \ref{eq:combinedDoverrd}),
  the galaxy data of
  \citet{Beutler11}, \citet{Ross15}, \citet{Alam17} and \citet{Bautista18}
  and the quasar data of \citet{Ata18} yields
\begin{equation}
\Omega_M=0.293 \pm 0.027 \hspace*{10mm} \Omega_\Lambda=0.675 \pm 0.099 
\end{equation}
corresponding to $\Omega_k=0.032 \pm 0.117$.
The best fit gives
$(c/H_0)/r_d=29.78 \pm0.55$ corresponding to
$hr_d=(0.683\pm 0.013)\times147.33$~Mpc.
The Pl2015 model has $\chi^2=13.76$ for 12
degrees of freedom and is within one standard deviation of the best fit,
as illustrated in Figure~\ref{figure::omol}.

\section{Conclusions}
\label{sec:conclu}

We have used Ly$\alpha$ and Ly$\beta$ spectral regions from the BOSS and eBOSS DR14 data sample to study BAO.
Following B17, we have built a model for the Ly$\alpha$ auto-correlation function
that we have then fit to the data. Our model incorporates the effects of redshift space distortions, the non-linear growth of matter, the contamination by metals and the modeling of high column density systems along  lines-of-sight to
quasars. 
Including  UV fluctuations has only a minor impact on the fit results.
We measure the ratios $D_H/r_d$ and $D_M/r_d$ at the average redshift of pixel pairs, $z=2.34$. We have also performed a measurement of these ratios from the Ly$\alpha$ auto-correlation function in two redshift bins, at $z=2.19$ and $z=2.49$.

The $D_H/r_d$ ratio is measured with a precision of $\sim 3.3$\%, a slight improvement over the precision obtained by B17 for this ratio.
The $D_M/r_d$ ratio is measured with a precision of $\sim 4.4$\%, which represents an improvement of about 25\% with respect to B17. The cosmological measurements obtained in this analysis are in agreement with the predictions of the flat $\Lambda$CDM model (Pl2015)
favored by the measurement of CMB anisotropies by Planck. 

We have also combined the measurements of the present analysis with the ones obtained from the cross-correlation of Ly$\alpha$ absorption and quasars by \citet{Blomqvist19}.
The latter alone favors a value of the $D_H/r_d$ ratio $\sim$ 3\% higher than the one favored by the Ly$\alpha$ auto-correlation.
As a result, the best-fit value of $D_H/r_d$ for the combined fit is shifted towards a higher value than the best-fit from the Ly$\alpha$ auto-correlation alone.
Combining the measurement of Ly$\alpha$ auto-correlation (this paper) with the quasar - Ly$\alpha$ cross-correlation of \citet{Blomqvist19}, the BAO measurements at $z=2.34$ are within $1.7\sigma$ of the predictions of the Pl2015 model.

The ensemble of BAO measurements is in good agreement with the
Pl2015 model
\citep{Planck16cosmo}.
They provide an independent way of determining cosmological parameters
that is based only on low-redshift measurements.
As illustrated in Fig. \ref{figure::omol}, the BAO results are also
consistent with the  recent Pantheon SNIa results \citep{Scolnic18}.

The present measurements will be much improved by the greater
statistical power of the upcoming DESI  \citep{DESI2016}
and WEAVE-QSO \citep{Weave16} projects.
The challenge will be to improve the physical modeling of the correlation
function in order to fully profit from the improved data.

\begin{acknowledgements}
We thank Pasquier Noterdaeme for providing the DLA catalog for eBOSS DR14 quasar.

Funding for the Sloan Digital Sky Survey IV has been provided by the Alfred P. Sloan Foundation, the U.S. Department of Energy Office of Science, and the Participating Institutions. SDSS acknowledges
support and resources from the Center for High-Performance Computing at the University of Utah. The SDSS web site is \url{http://www.sdss.org/}.

MB, MP and IPR were supported by the A*MIDEX project (ANR-11-IDEX-0001-02) funded by the ``Investissements d'Avenir'' French Government program, managed by the French National Research Agency (ANR), and by ANR under contract ANR-14-ACHN-0021

SDSS is managed by the Astrophysical Research Consortium for the Participating Institutions of the SDSS Collaboration including the Brazilian Participation Group, the Carnegie Institution for Science, Carnegie Mellon University, the Chilean Participation Group, the French Participation Group, Harvard-Smithsonian Center for Astrophysics, Instituto de Astrofísica de Canarias, The Johns Hopkins University, Kavli Institute for the Physics and Mathematics of the Universe (IPMU) / University of Tokyo, Lawrence Berkeley National Laboratory, Leibniz Institut für Astrophysik Potsdam (AIP), Max-Planck-Institut für Astronomie (MPIA Heidelberg), Max-Planck-Institut für Astrophysik (MPA Garching), Max-Planck-Institut für Extraterrestrische Physik (MPE), National Astronomical Observatories of China, New Mexico State University, New York University, University of Notre Dame, Observatório Nacional / MCTI, The Ohio State University, Pennsylvania State University, Shanghai Astronomical Observatory, United Kingdom Participation Group, Universidad Nacional Autónoma de México, University of Arizona, University of Colorado Boulder, University of Oxford, University of Portsmouth, University of Utah, University of Virginia, University of Washington, University of Wisconsin, Vanderbilt University, and Yale University.

\end{acknowledgements}

\bibliographystyle{aa}
\bibliography{bibi_article}

\begin{appendix}

\begin{figure*}
	\includegraphics[width=.5\textwidth]{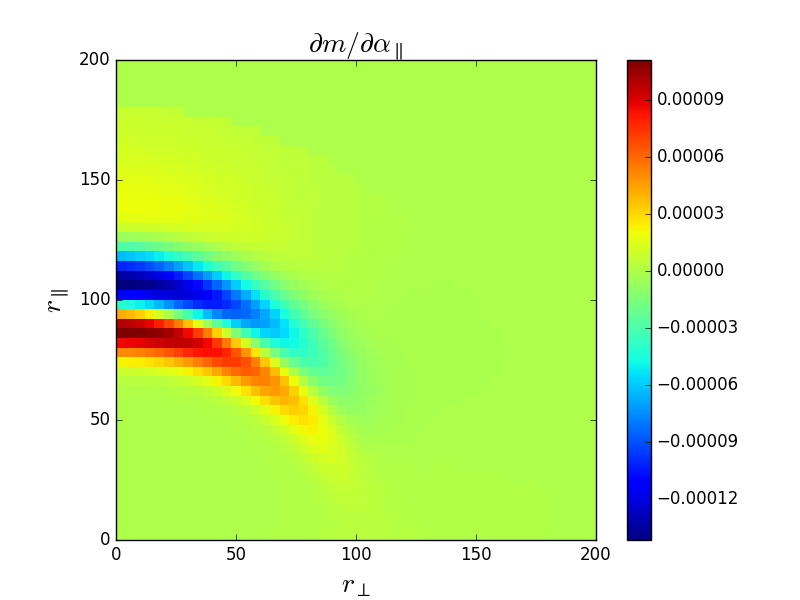}
         \includegraphics[width=.5\textwidth]{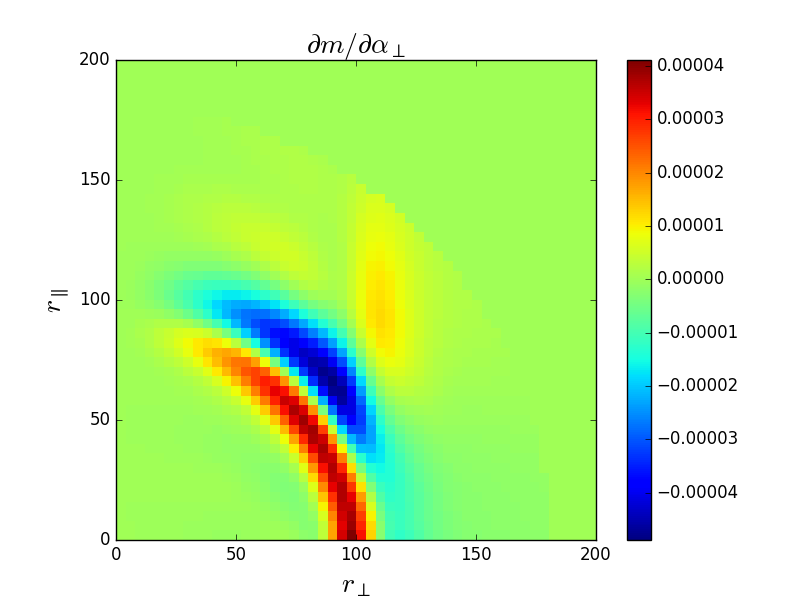}\\
	\includegraphics[width=.5\textwidth]{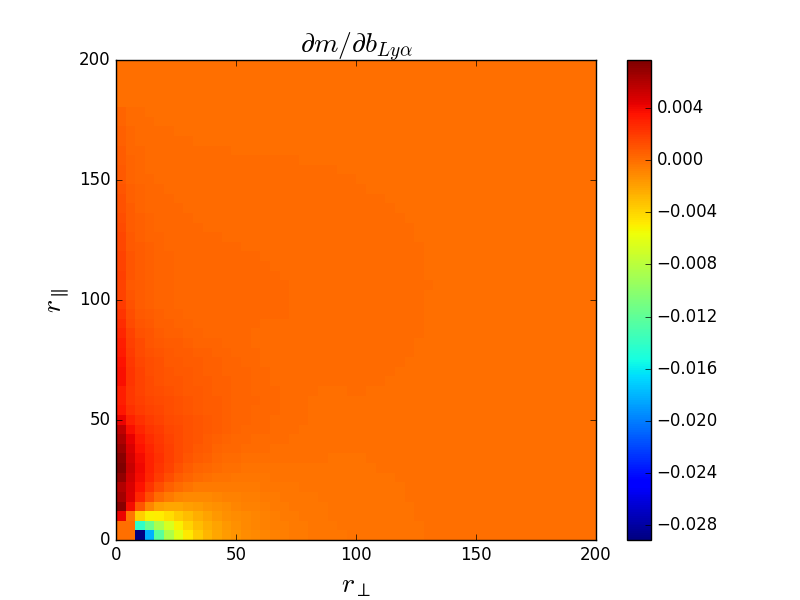}
        \includegraphics[width=.5\textwidth]{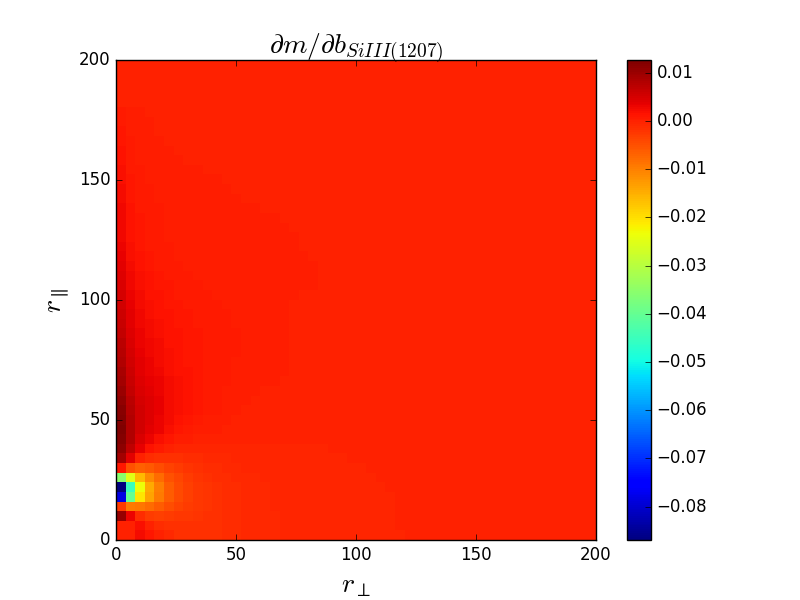}
	\caption{\label{fig:effz}
          The quantity $\partial m/\partial p$  in the ($r_\perp$, $r_\parallel$)
          plane for the fitted parameters $p \in [\alpha_{\parallel}, \alpha_{\perp},\blya ,b_{\rm SiIII(1207)}]$. $m$ is the baseline model of the \lya auto-correlation function. 
The graphs show which pixels contribute the most to the constraints on the considered parameter.
BAO parameters $\alpha_{\parallel}$ and $\alpha_{\perp}$ are constrained by the bins around the location of the BAO peak, while the \lya bias is mostly constrained by the bins at $\sim$ zero separation. 
We also note that the \ion{Si}{III}(1207) bias is mostly constrained by the bins $r_{\perp}\sim 0$, $r_{\parallel}\sim 21$\hMpc, in agreement with the $r^{ap}_\parallel$ apparent separation given in Table \ref{table:metal_auto}.
}
\end{figure*}

\section{Effective redshift of the fitted parameters}
\label{appendixA}

  In this section we present a method to determine
  the region in $(\rperp,\rpar,z)$ space that
is most constraining for the various parameters in the fits of the
correlation function.
We can expect that the parameters $(\aperp,\apar)$ are mostly determined
by  $(\rperp,\rpar)$ bins near the BAO peak and at a redshift
near  the mean redshift of the pixel pairs used in the BAO region.
In fact, 
previous studies (B17, \citet{Busca13, Delubac15})
defined the effective redshift of the BAO measurement in this way.
Here, we make this intuitive conclusion more precise by using
a Fisher matrix analysis.

We use the Fisher matrix formalism as follows: given a parameter $p$ varying linearly with redshift, we define the effective redshift $z_0$ at which it is measured by:
\begin{equation}
  \label{eq:zdep}
  p(z) = p_0 + p_1(z-z_0),
\end{equation}
where $p_0$ is the value given by the fit at $z=z_0$. 
The covariance matrix $C_p$ between two parameters $p_0$ and $p_1$ is given by: 
\begin{equation}
  C_{p} \equiv \begin{pmatrix}
    \sigma_0^2 & \rho \sigma_0 \sigma_1  \\ \rho \sigma_0 \sigma_1 & \sigma_1^2 
  \end{pmatrix},
\end{equation}
where $\sigma_i^2$ is the variance of the parameter $p_i$ and $\rho$ is the correlation coefficient between $p_0$ and $p_1$.
By definition, $C_p$ is the inverse of the Fisher matrix $F_p$ : 
\begin{equation}
  C_{p}^{-1} \equiv 
  F_{p} \equiv  \begin{pmatrix}
    \sum_{ij} \frac{\partial m_i}{\partial p_0}C_{ij}^{-1} \frac{\partial m_j}{\partial p_0} &\sum_{ij} \frac{\partial m_i}{\partial p_0}C_{ij}^{-1} \frac{\partial m_j}{\partial p_1} \\
    \sum_{ij} \frac{\partial m_i}{\partial p_1}C_{ij}^{-1} \frac{\partial m_j}{\partial p_0} & \sum_{ij} \frac{\partial m_i}{\partial p_1}C_{ij}^{-1} \frac{\partial m_j}{\partial p_1}
  \end{pmatrix},
\end{equation} 
with $m_i$ the model at bin $i$. In the case of the linear redshift dependency (\ref{eq:zdep}), the Fisher matrix $F_p$ at redshift $z$, computed using the set of all the fitted parameter values, $\{\lambda_0\}$,  is given by : 
\begin{equation}
  F_p(z) = \sum_{i,j} \frac{\partial m_i}{\partial p} \Big|_{\{\lambda_0\}}C_{ij}^{-1} \frac{\partial m_j}{\partial p}\Big|_{\{\lambda_0\}}
  \begin{pmatrix}
    1 & (z_j-z) \\
    (z_i-z) & (z_i-z)(z_j-z)
  \end{pmatrix},
\end{equation}
with $z_i$ the mean redshift of the pairs in bin $i$. \\
We represent the quantities $\frac{\partial m_i}{\partial p}$ for 4 of the 12 fitted parameters of the \lya auto-correlation function in Fig.\ref{fig:effz}. 
The covariance matrix $C_p(z)$ then reads : 
\begin{equation}
  C_p(z) = \frac{1}{|F_p|} \sum_{ij} M_{ij}
  \begin{pmatrix}
    (z_i-z)(z_j-z) & -(z_i-z) \\
    -(z_j-z) & 1
  \end{pmatrix}, 
\end{equation}	
with 
\begin{equation}
  M_{ij} \equiv \frac{\partial m_i}{\partial p}\Big |_ {\{\lambda_0\}} C^{-1}_{ij} \frac{\partial m_j}{\partial p}\Big |_ {\{\lambda_0\}}.
\end{equation}
 
Since $M_{ij}$ is symmetric, the determinant of the Fisher matrix, $|F_{p}|$, does not depend on redshift and is given by:
\begin{equation}
  |F_{p}| = \sum_{i,j,k,l} M_{ij} M_{kl} z_i\times (z_j-z_k).
\end{equation}	

The variance of $p_0$ at redshift $z$ becomes: 
\begin{equation}
  \sigma^2_0(z) = \frac{1}{|F_p|} \sum_{ij} M_{ij}(z_i-z)(z_j-z).
\end{equation}
The effective redshift $z_0$ is the value which minimizes the error on $p_0$ : 
\begin{equation}
  \frac{d\sigma_0^2}{dz} \Big |_{z_0} = - \frac{2}{|F_p|} \sum_{ij} M_{ij}(z_i-z_0) = 0,
\end{equation}
i.e., 
\begin{equation}
  z_0 = \frac{\sum_{ij} M_{ij}z_i}{\sum_{ij}M_{ij}}.
\end{equation}
In the case of a combined fit, we compute one matrix $M^d$ for each correlation  function entering the fit: 
\begin{equation}
  M_{ij}^d \equiv \frac{\partial m^d_i}{\partial p}\Big |_ {\{\lambda_0\}} (C^{d}_{ij})^{-1} \frac{\partial m^d_j}{\partial p}\Big |_ {\{\lambda_0\}}, 
\end{equation}	
where $m^d_i$ is the model for the correlation function $d$ at bin $i$.
In this case, $z_0$ reads: 
\begin{equation}
\label{eq:z0}
  z_0 = \frac{\sum_d \sum_{ij} M_{ij}^d z_i}{\sum_d \sum_{ij}M^d_{ij}}.
\end{equation}

Table \ref{table:z_effective} presents the effective redshifts at which the $\alpha_\parallel$ and $\alpha_\perp$ parameters are  measured for the different correlation functions computed in this paper. The effective redshift values differ by less than 0.5\% for $\alpha_\parallel$ and  $\alpha_\perp$. Figure \ref{fig:effz} shows the quantities $\partial m/\partial p$ in the ($r_\perp$, $r_\parallel$) plane for the fitted parameters $p \in [\alpha_{\parallel}, \alpha_{\perp},\blya ,b_{\rm SiIII(1207)}]$. $m$ is the baseline model for the Ly$\alpha$ auto-correlation.

\begin{table}
	\caption{\label{table:z_effective} Effective redshifts at which the $\alpha_\parallel$ and $\alpha_\perp$ parameters are measured. The average redshift of pairs is also given. \autolya (low z) and (high z) are introduced in Appendix~\ref{sec:tomo}.} 	
	\begin{tabular}{cccc}
	\hline 
	\hline 
	Correlation functions & $\overline{z}$ & $z_{\alpha_{||}}$ & $z_{\alpha_{\perp}}$\\ 
	\hline 
        \\
        \autolya & 2.35 & 2.34 & 2.34 \\
        \crosslya & 2.29 & 2.29 & 2.28 \\
        Ly$\alpha$(Ly$\alpha$)xLy$\alpha$(Ly$\alpha$+Ly$\beta$)~ & 2.34 & 2.33 & 2.33 \\
        \hline
        \autolya (low z) & 2.19 & 2.19 & 2.18 \\
        \autolya (high z) & 2.49 & 2.49 & 2.49 \\
        \crosslyb & 2.76 & 2.77  & 2.78 \\       
          \hline 
	\end{tabular} 
\end{table}

\section{Fits in two redshift bins}
\label{sec:tomo}

\begin{figure}
\includegraphics[width=\columnwidth]{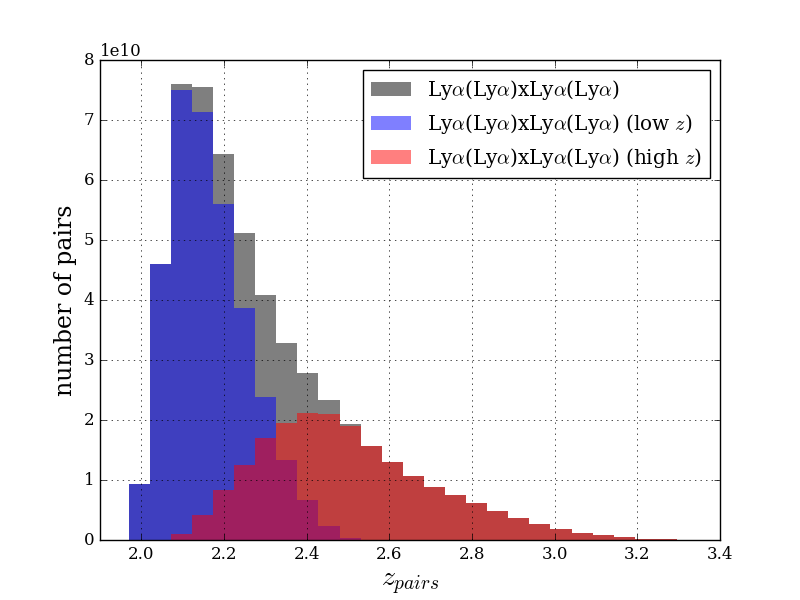}
	\caption{Pixel pair redshift distribution of the subsamples used in the present analysis: full Ly$\alpha$ auto-correlation function (gray), low-redshift Ly$\alpha$ auto-correlation function (blue), high redshift Ly$\alpha$ auto-correlation function (red). The two latter subsamples are used to produce a measurement of $H(z)$ at $z=2.19$ and $z=2.49$.
	}
        \label{fig:histo_pairs}
\end{figure}

The present data set is large enough to constrain the BAO parameters in two independent redshift bins, in a way similar to what will be done in forthcoming cosmological surveys, such as the DESI project \citep{DESI2016}.
To simplify the analysis, we consider only the \autolya correlation function.

A straightforward way of defining a high- and  low-redshift sample of pixel pairs would be to simply use pixel pairs of mean redshift less than or greater than an appropriately chosen value, $z_{\rm cut}$. The drawback of such an approach is that a given pair of {\it forests} could belong to both bins, as some pixels in a given forest would be associated with some pixels in the other forest, either in pixel pairs with mean redshift less than $z_{\rm cut}$, or in pixel pairs with mean redshift greater than $z_{\rm cut}$. The fact that some pairs of forests belong to both redshift bins introduces unwanted correlations when correcting for the distortions introduced by our continuum fitting procedure. To circumvent this problem, we choose to assign forest pairs to the high or low-redshift sample by cutting on the mean of the maximum $z$ of the two forests.

We thus evaluate, for all pairs of forests ($i,j$), the following quantity:
\begin{equation}
z_{ij}=\frac{z_{\rm max,abs}^i+z_{\rm max,abs}^j}{2},
\end{equation}
where $1+z_{\rm max,abs}^k = {\rm max}(\lambda_{\rm obs}^k)/\lambda_{\rm abs}$, with ${\rm max}~\lambda_{\rm obs}^k$ the last pixel of forest $k$, and $\lambda_{\rm abs}$ the rest-frame wavelength of the considered transition. The condition $z_{ij}<z_{\rm cut}$ defines the low redshift bin, while the opposite condition defines the high redshift one. 

The value of $z_{\rm cut}$ is tuned so that the sum of the weights of all absorber pairs in the high redshift correlation function equals the one of all pairs in the low redshift correlation function. This ensures the two bins have a comparable statistical power.
This process leads us to select $z_{\rm cut}=2.5$.
The redshift distribution of absorber pairs obtained in this way are shown on Figure \ref{fig:histo_pairs}. The average pair redshift is $\overline{z}=2.19$ and $\overline{z}=2.49$ for the low and high redshift bin, respectively.

\begin{figure}
\includegraphics[width=\columnwidth]{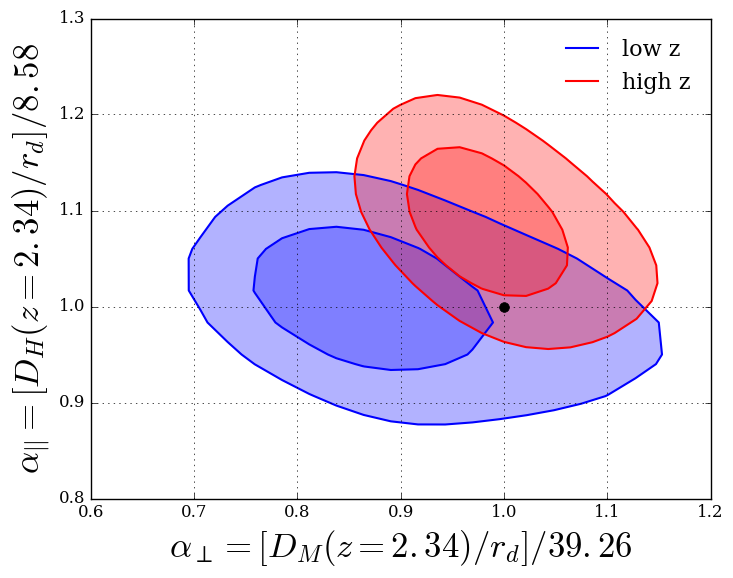}
\caption{The 68\% and 95\% confidence level contours
  in the ($\alpha_\parallel$,$\alpha_\perp$) plane from
  the \autolya computed with the low and high redshift bins.
  The $\Delta\chi^2$ values corresponding to confidence levels
  are taken from Table \ref{table::confidence}.
  The black dot corresponds to the Pl2015 model.
}
	\label{fig:contours_binsz}
\end{figure}

\begin{figure*}
\centerline{\includegraphics[width=\textwidth]{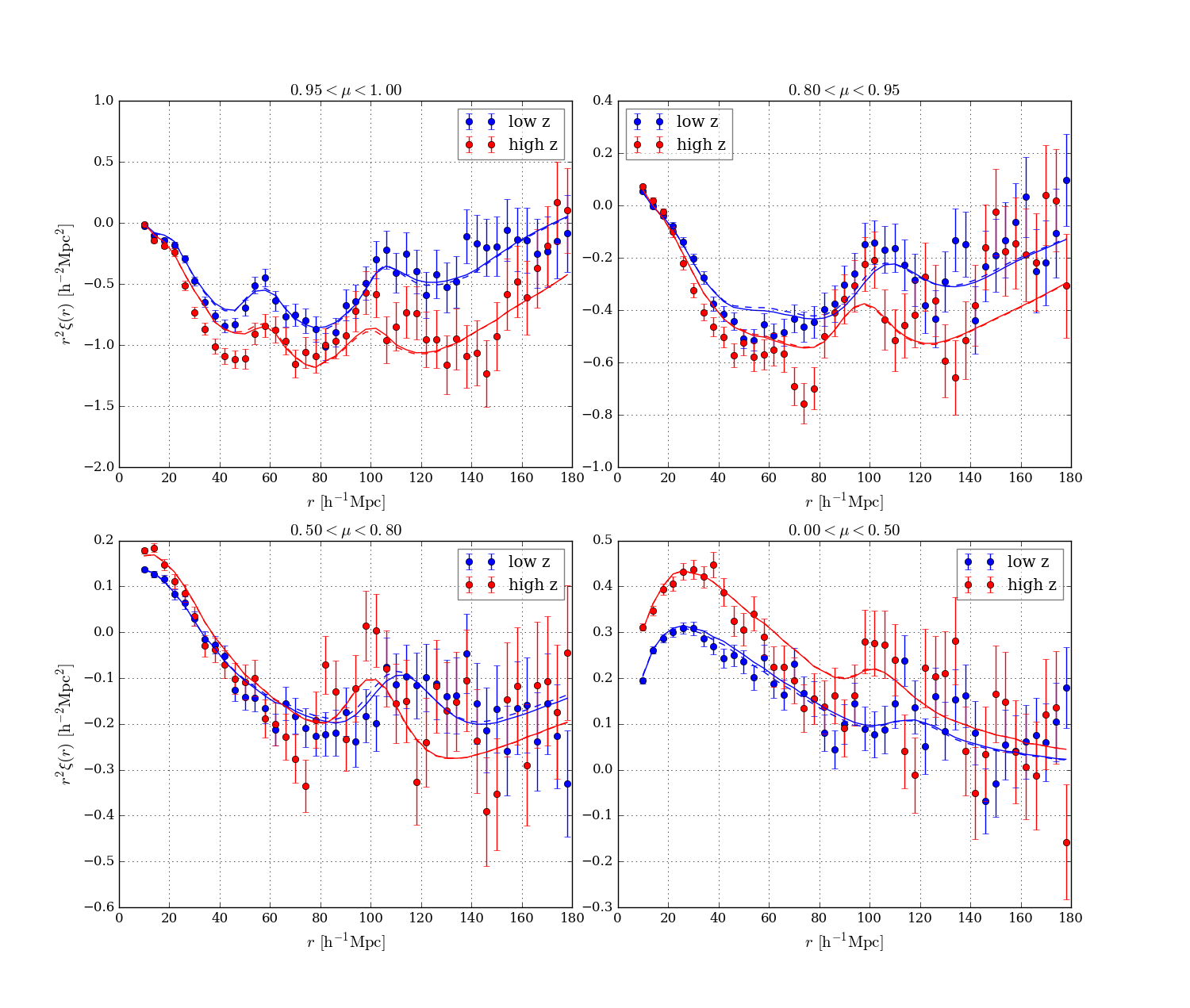}}
	\caption{\label{fig:cf_bins} Ly$\alpha$(Ly$\alpha$)xLy$\alpha$(Ly$\alpha$) function in four ranges of $\mu=r_\parallel/r$ values, computed in a low redshift (blue dots) and in a high redshift (red dots) bin obtained by splitting our sample so that the two bins have equivalent statistical power. The dashed lines correspond to the simple fits to the data of the Ly$\alpha$(Ly$\alpha$)xLy$\alpha$(Ly$\alpha$) correlation function, the solid line, to the combining fits to the data of the Ly$\alpha$(Ly$\alpha$)xLy$\alpha$(Ly$\alpha$) and  Ly$\alpha$(Ly$\alpha$)xLy$\alpha$(Ly$\beta$) correlation functions. }
\end{figure*}

\begin{table}
  \centering
  \caption{
      Results in two redshift bins from fitting the baseline model to the  \autolya correlation function. \label{tab:splitz}}
  \begin{tabular}{ccc}
    \hline
    \hline
    Parameters           & low z  & high z  \\
    \hline
    $\alpha_{\parallel}$ & 1.008 $\pm$ 0.043 &  1.088 $\pm$ 0.046  \\
    $\alpha_{\perp}$ & 0.861 $\pm$ 0.062 &  0.977 $\pm$ 0.044  \\ 
    \hline
    $\betalya$ & 2.083 $\pm$ 0.160  & 1.585 $\pm$ 0.113  \\
    $b_{\eta Ly\alpha}$&-0.218 $\pm$ 0.006  &  -0.201 $\pm$ 0.005  \\
    \hline
    $\betahcd$ & 0.745 $\pm$ 0.174  & 0.678 $\pm$ 0.179 \\
    $\bhcd$ & -0.058 $\pm$ 0.007  & -0.040 $\pm$ 0.006 \\
    \hline
    $b_{\rm SiII(1190)}$ & -0.0049 $\pm$ 0.0015 &  -0.0045 $\pm$ 0.0016 \\
    $b_{\rm SiII(1193)}$ & -0.0067 $\pm$ 0.0015 &  -0.0021 $\pm$ 0.0015 \\
    $b_{\rm SiIII(1207)}$ & -0.0105 $\pm$ 0.0016 &  -0.0055 $\pm$ 0.0017 \\
    $b_{\rm SiII(1260)}$ &  -0.0039 $\pm$ 0.0018 &  -0.0021 $\pm$ 0.0017  \\
    $b_{\rm CIV(eff)}$ & -0.0178 $\pm$ 0.0095 &  -0.0186 $\pm$ 0.0095 \\
    \hline
    $\chi^2_{\rm min}$ & 1580.95 &  1737.15 \\
    $DOF$ & 1590 - 11  & 1590 - 11  \\
    Probability & 0.481 & 0.003 \\
    $\chi^2(\alpha_{\parallel}=\alpha_{\perp}=1)$ & 1584.30 & 1740.89 \\ 
    \hline 
  \end{tabular}
  \centering
\end{table}

Figure \ref{fig:cf_bins} presents the result of fitting the \autolya correlation baseline model
to the data in the low (blue points) and high (red points) redshift bins
in the usual four $\mu$ wedges.

Table \ref{tab:splitz} shows the associated best-fit parameters. 
From the table, we note that $\beta_{\rm Ly\alpha}$ is notably different at low and high redshift. When fitting the full sample, we assumed, following \citet{Kirkby13}, $\beta_{\rm Ly\alpha}$ to be constant. A redshift dependent $\beta_{\rm Ly\alpha}$ could thus be an improvement in future analyses. 

From the lower right panel  of Figure \ref{fig:cf_bins}, we see that the amplitude of the high-redshift correlation function is higher than the amplitude of the low-redshift one. This is expected, as $b_{\rm Ly\alpha}$  increases with redshift \citep{Kirkby13}.

Figure \ref{fig:contours_binsz} presents the constraints obtained, in the ($\alpha_\perp$,$\alpha_\parallel$) parameter space, from fitting the Ly$\alpha$ auto-correlation Ly$\alpha$(Ly$\alpha$)xLy$\alpha$(Ly$\alpha$) in the low (blue contours) and high (red contours) redshift bins.
  The values of $\Delta\chi^2$ corresponding to a given confidence level
  were taken from Table \ref{table::confidence}.
  Both the high and low redshift measurements are within
$1.2\sigma$ of the Pl1015 model.

\section{The \lya(\lya)x\lyb(\lyb) cross correlation}
\label{sec:lyblyb}

\begin{table}
  \centering
  \caption{Same as Table \ref{table:metal_auto} for the main metal/metal, metal/\lyb and \lya/metal correlations relevant to the computation of the \crosslyb correlation function. The apparent separation $r_{||}^{ap}=(1+\overline{z})H(\overline{z})\Big (\frac{\lambda_1}{\lambda_\alpha}-\frac{\lambda_2}{\lambda_\beta} \Big)$ is computed at the average redshift of 2.76.}
  \label{table:metal_cross}
  \begin{tabular}{ccc}       
    \hline
    \hline
    Transitions & $\lambda_1/\lambda_2$ & $r_{||}^{ap}$[\hMpc]\\
    \hline
    \ion{Si}{II}(1190)/\ion{O}{VI}(1038) & 1.147 & -88 \\
    \ion{Si}{II}(1193)/\ion{O}{VI}(1038) & 1.150 & -81 \\
    \ion{Si}{II}(1190)/\ion{O}{VI}(1032) & 1.154 & -73 \\
    \ion{Si}{II}(1193)/\ion{O}{VI}(1032) & 1.156 & -66 \\
    \ion{Si}{II}(1190)/\lyb(1026) & 1.161 & -56 \\
    \ion{Si}{III}(1207)/\ion{O}{VI}(1038) & 1.163 & -52 \\
    \ion{Si}{II}(1193)/\lyb(1026) & 1.163 & -50 \\
    \ion{Si}{III}(1207)/\ion{O}{VI}(1032) & 1.169 & -37 \\
    \lya(1216)/\ion{O}{VI}(1038) & 1.172 & -31 \\
    \ion{Si}{III}(1207)/\lyb(1026) & 1.176 & -20 \\
    \lya(1216)/\ion{O}{VI}(1032) & 1.178 & -16 \\
    \ion{Si}{II}(1260)/\ion{O}{VI}(1038) & 1.215 & 68 \\
    \ion{Si}{II}(1260)/\ion{O}{VI}(1032) & 1.221 & 83 \\
    \ion{Si}{II}(1260)/\lyb(1026) & 1.229 & 100 \\
    \hline        
  \end{tabular}
\end{table}

\begin{figure}
	\includegraphics[width=\columnwidth]{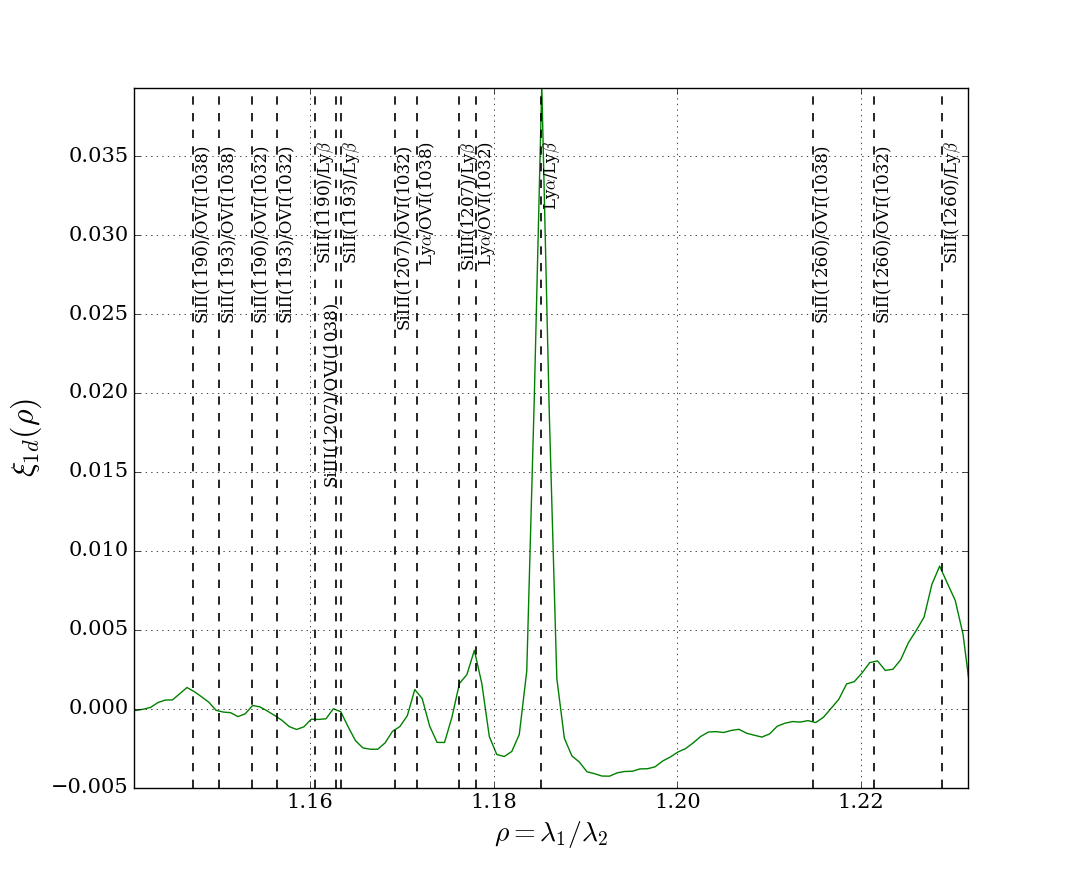}
	\caption{\label{fig:c1d_cross} Same as Fig. \ref{fig:c1d_auto} for cross-correlation function of Ly$\alpha$ with Ly$\beta$ regions, as a function of the ratio of transition wavelengths.}
\end{figure} 

\begin{table}
  \caption{\label{tab:result_cross}Results of the combining fit on \autolya, \crosslya and \crosslyb correlation functions with the BAO parameters $(\alpha_{\parallel},\alpha_{\perp})$ fixed to 1.}
  \begin{tabular}{cccc}
    \hline
    \hline
    Parameters & Ly$\alpha$(Ly$\alpha$)xLy$\alpha$(Ly$\alpha$) \\
               & + Ly$\alpha$(Ly$\alpha$)xLy$\alpha$(Ly$\beta$) \\  &  + Ly$\alpha$(Ly$\alpha$)xLy$\beta$(Ly$\beta$)\\
    \hline
    $\betalya$ &  1.840 $\pm$  0.084\\
    $b_{\eta Ly\alpha}$& -0.212 $\pm$  0.004\\
    \hline
    $\betalyb$ & 1.123 $\pm$  0.384\\
    $b_{\eta Ly\beta}$ & -0.098 $\pm$  0.018\\
    \hline
    $\betahcd$ & 1.116 $\pm$  0.152 \\
    $\bhcd^{\rm Ly \alpha(Ly \alpha) \times Ly \alpha(Ly \alpha)}$ & -0.050 $\pm$  0.004\\
    $\bhcd^{\rm Ly \alpha(Ly \alpha) \times Ly \alpha(Ly \beta)}$ & -0.073 $\pm$  0.005\\
    $\bhcd^{\rm Ly \alpha(Ly \alpha) \times Ly \beta(Ly \beta)}$ & -0.002 $\pm$  0.032\\
    \hline
    $b_{\rm OVI(1032)}$ & -0.0081 $\pm$  0.0015\\
    $b_{\rm OVI(1038)}$ & -0.0055 $\pm$  0.0014\\
    $b_{\rm SiII(1190)}$ & -0.0025 $\pm$  0.0005\\
    $b_{\rm SiII(1193)}$ & -0.0022 $\pm$  0.0005\\
    $b_{\rm SiIII(1207)}$ & -0.0035 $\pm$  0.0005\\
    $b_{\rm SiII(1260)}$ & -0.0011 $\pm$  0.0006\\
    $b_{\rm CIV(eff)}$ & -0.0047 $\pm$  0.0025\\
    \hline
    $\chi^2_{\rm min}/DOF$ &6469.77/(6360-15) \\
    Probability &0.134 \\
    \hline 
  \end{tabular}
  \centering
\end{table}

\begin{figure*}
\centerline{\includegraphics[width=\textwidth]{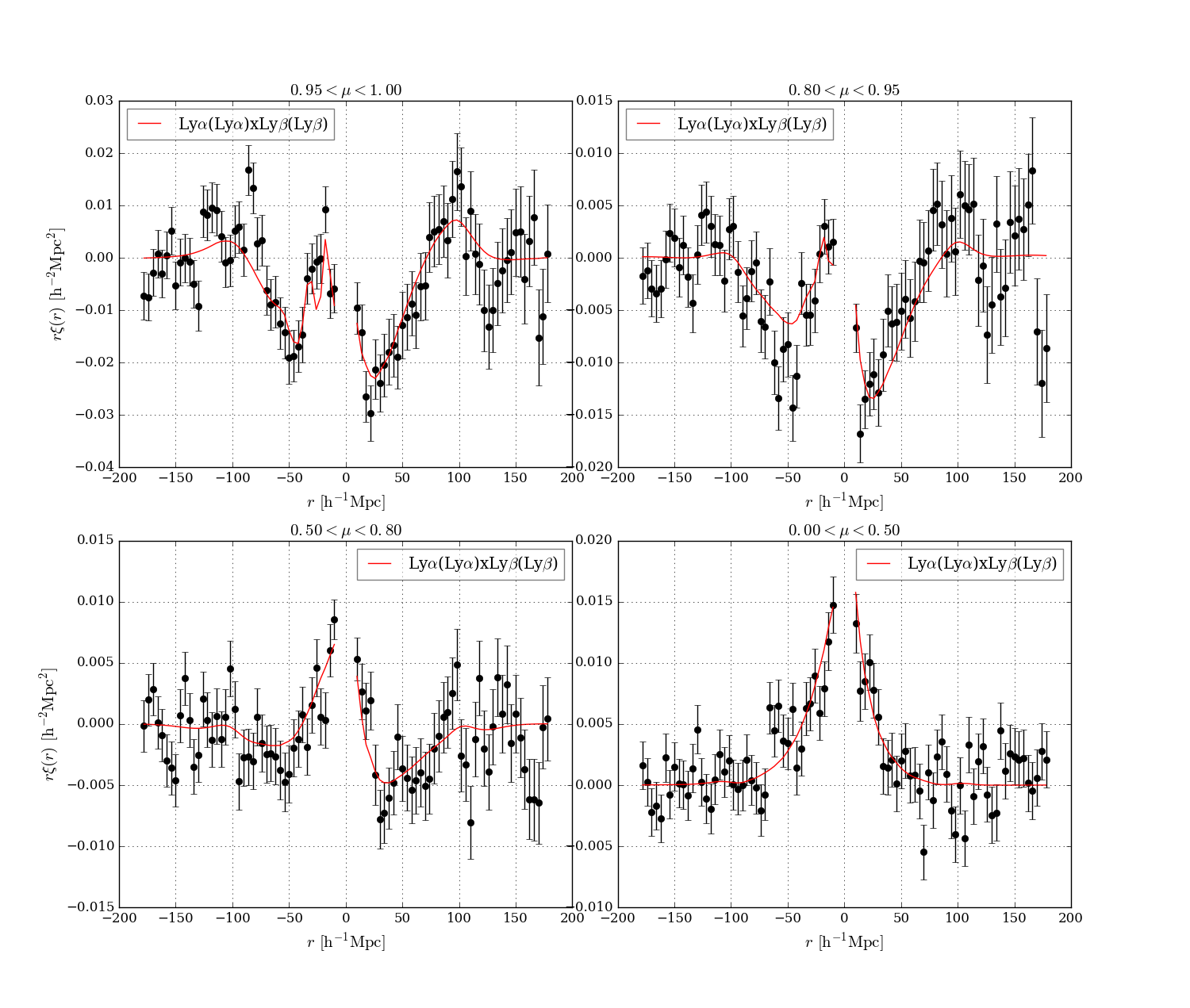}}
\caption{\label{fig:lya_lyaxlyb_lyb}Measured \crosslyb correlation function in four ranges of $\mu$. In order to see the asymmetry of this correlation function, $r =(\rpar^2+\rperp^2)^{1/2}$ is multiplied by the sign of
  $\rpar$ which is positive if the \lya absorber is farther than the \lyb absorber, and negative in the opposite configuration. The model for this correlation function (red solid line) is fitted on \autolya, \crosslya and \crosslyb.}
\end{figure*}

As an extension of our main analysis, we compute the \crosslyb correlation function, following the procedure previously described. We computed the 1D correlation function in the \lyb region (Fig \ref{fig:c1d_cross}) to identify the contaminating metals (see Table \ref{table:metal_cross}).

The results of this analysis are presented in Table \ref{tab:result_cross}. 
The \lyb absorption signal is clearly detected as $b_{\rm Ly\beta}$ is non zero at at the 2.9 $\sigma$ level. 
We also see the signal due to \ion{O}{vi}(1032)  and \ion{O}{vi}(1036).
Note that the correlation between \lyb and \ion{Si}{ii}(1260) occurs
near the BAO peak (last line of Table \ref{table:metal_cross}). Due to the small \lyb absorption cross-section and to the small wavelength extent of the \lyb region, it is harder to detect the BAO peak than for the \lya auto-correlation function.  Moreover, the \lyb-\ion{Si}{ii} correlation further overlaps the BAO signal at small $\rperp$, hampering its detection in our data.

Figure \ref{fig:lya_lyaxlyb_lyb} presents for the first time the 2D \crosslyb correlation function. It is shown in the usual four wedges of $\mu$ values, as a function of $r = \sqrt{r_{\parallel}^2+r_{\perp}^2}$ multiplied by the sign of $r_{\parallel}$. Note that the model is not symmetric around zero separation.

  The oscillator strength of \lyb absorption is a fifth of that of \lya,
  and consequently there are far fewer \lyb  HCD systems
  than \lya  HCD systems.
  On the other hand \lyb  absorption in our analysis occurs at a
  systematically higher redshift.
  Overall we find that
  $b_{\rm HCD}^{\rm Ly\alpha(Ly\alpha)\times Ly\beta(Ly\beta)}$
  is consistent with zero.

In summary, there are not enough data at present to constrain the BAO peak position with \lyb absorption only. However, \lyb absorption could be used to access to physical IGM parameters at redshifts for which the \lya absorption is saturated (\citet{Dijkstra04}, \citet{Irsic14}).

\section{Confidence levels}
\label{sec:cl}
		
To make a precise estimate of the relation between
$\Delta\chi^2$ and confidence level,
we followed closely the procedure of \citet{Bourboux17}.
We
generated a large number of  simulated correlation functions using the fiducial
cosmological model and the best-fit values of non-BAO parameters, 
randomized using the covariance matrix measured with the data.
Each simulated correlation function was then fit for the model parameters
and the $\chi^2$ for the best-fit parameters
compared with the best $\chi^2$ with
one or more parameters set to the known input values.
Confidence levels are the fractions of the generated data sets
that have best fits below the $\Delta\chi^{2}$ limit.
The uncertainties are  estimated using a bootstrap technique.

  The analysis of \citet{Bourboux17}
  followed this procedure using  models that
incorporated only \lya absorption and models that incorporated
also HCDs and metals.
Since no significant differences were seen in the two methods,
we use here only \lya absorption, considerable simplifying
the analysis.

The results are summarized 
in Table \ref{table::confidence} for various correlation functions.
In all cases the $\Delta\chi^2$ values corresponding to a given
confidence level are increased above the standard values.
For example, for the \alllya~correlation, the one- and two-standard
deviation contours for $(\apar,\aperp)$ correspond to
$\Delta\chi^2=2.77$ and $\Delta\chi^2=7.33$ to be compared with
the standard values of 2.29 and 6.18.

\begin{table}[tb]
\centering
\caption{Values of $\Delta\chi^{2}$ corresponding to confidence levels (CLs) $(68.27,95.45\%)$.
  Values are derived from 10,000 Monte Carlo simulations of the correlation function that are
  fit using the model containing only Ly$\alpha$ absorption.
}
\label{table::confidence}
\begin{tabular}{l c c}
\hline
\hline
\noalign{\smallskip}
Parameter & $\Delta\chi^{2}$ $(68.27\%)$ & $\Delta\chi^{2}$ $(95.45\%)$ \\
\noalign{\smallskip}
\hline
\noalign{\smallskip}
\alllya & & \\
$\alpha_{\parallel}$ & 1.19 $\pm$ 0.03 & 4.74 $\pm$ 0.09\\
$\alpha_{\perp}$& 1.23 $\pm$ 0.03 & 4.83 $\pm$ 0.08\\
$(\alpha_{\parallel},\alpha_{\perp})$ &2.77 $\pm$ 0.04 & 7.33 $\pm$ 0.10\\
\noalign{\smallskip}
\hline
\noalign{\smallskip}
\alllya & & \\
+ QSO $\times$ \lyalyab & & \\
$\alpha_{\parallel}$ & 1.08$\pm$0.02 & 4.29$\pm$0.10 \\
$\alpha_{\perp}$ & 1.08$\pm$0.02 & 4.28$\pm$0.10 \\
$(\alpha_{\parallel},\alpha_{\perp})$ &2.47$\pm$0.03 & 6.71$\pm$0.13 \\
\noalign{\smallskip}
\hline
\noalign{\smallskip}
\autolya  & & \\
$\alpha_{\parallel}$ & 1.19 $\pm$ 0.02 & 4.65 $\pm$ 0.09\\
$\alpha_{\perp}$& 1.17 $\pm$ 0.02 & 4.32 $\pm$ 0.07\\
$(\alpha_{\parallel},\alpha_{\perp})$ & 2.65 $\pm$ 0.04 & 6.99 $\pm$ 0.10\\
\noalign{\smallskip}
\hline
\noalign{\smallskip}
low z  & & \\
$\alpha_{\parallel}$ &  1.28 $\pm$ 0.02 & 5.09 $\pm$ 0.08\\
$\alpha_{\perp}$ &  1.35 $\pm$ 0.02 & 5.09 $\pm$ 0.10\\
$(\alpha_{\parallel},\alpha_{\perp})$ &  2.89 $\pm$ 0.04 & 7.55 $\pm$ 0.12\\
\noalign{\smallskip}
\hline
\noalign{\smallskip}
high z  & & \\
$\alpha_{\parallel}$ &  1.28 $\pm$ 0.03 & 4.92 $\pm$ 0.09\\
$\alpha_{\perp}$ &  1.23 $\pm$ 0.03 & 4.74 $\pm$ 0.08\\
$(\alpha_{\parallel},\alpha_{\perp})$ &  2.83 $\pm$ 0.04 & 7.44 $\pm$ 0.11\\
\noalign{\smallskip}
\hline
\end{tabular}
\end{table}

\end{appendix}

\end{document}